\documentstyle[preprint,tighten,eqsecnum,aps,floats,epsfig]{revtex}

\def\ub{{\overline{u}}}
\def\vb{{\overline{v}}}

\begin{document}
\draft

\title{
The $N$-component Ginzburg-Landau Hamiltonian with cubic anisotropy: a six-loop study.
}
\author{Jos\'e Manuel Carmona$\,^1$, Andrea Pelissetto$\,^2$, and 
Ettore Vicari$\,^1$ }
\address{$^1$
Dipartimento di Fisica dell'Universit\`a 
and I.N.F.N., 
Via Buonarroti 2, I-56127 Pisa, Italy.}
\address{$^2$ Dipartimento di Fisica dell'Universit\`a di Roma I
and I.N.F.N., I-00185 Roma, Italy \\
{\bf e-mail: \rm
{\tt carmona@mailbox.difi.unipi.it}, 
{\tt Andrea.Pelissetto@roma1.infn.it},
{\tt vicari@mailbox.difi.unipi.it}
}}

\date{\today}

\maketitle

\begin{abstract}
We consider the Ginzburg-Landau Hamiltonian with a 
cubic-symmetric quartic interaction 
and compute the renormalization-group functions to six-loop order in 
$d=3$. We analyze the stability of the fixed points using a 
Borel transformation and a conformal mapping that takes into account 
the singularities of the Borel transform. 
We find that the cubic fixed point is stable for $N>N_c$, $N_c = 2.89(4)$. 
Therefore, the critical 
properties of cubic ferromagnets are not described by the Heisenberg
isotropic Hamiltonian, but instead by the cubic model at the cubic fixed point.
For $N=3$, the critical exponents at the cubic and symmetric fixed points differ 
very little (less than the precision of our results, which is 
$\lesssim 1\%$ in the case of $\gamma$ and $\nu$). Moreover,
 the irrelevant interaction bringing from the symmetric to the cubic fixed point gives 
rise to slowly-decaying scaling corrections with exponent $\omega_2=0.010(4)$.
For $N=2$, the isotropic fixed point is stable and the cubic interaction
induces scaling corrections with exponent $\omega_2 = 0.103(8)$. These
conclusions are confirmed by a similar analysis of the five-loop 
$\epsilon$-expansion. A constrained analysis which takes into account that 
$N_c = 2$ in two dimensions gives $N_c = 2.87(5)$.
\end{abstract}

%\vskip1.5pc
%\bgroup\small
%\leftskip=0.10753\textwidth \rightskip\leftskip
%\noindent{\bf Keywords:} 
%Critical Phenomena, Critical Exponents, Ferromagnetic Materials, Cubic Anisotropy,
%Renormalization Group. 
%\par\egroup
%\vskip1.5pc

\pacs{PACS Numbers: 75.10.Hk, 05.70.Jk, 64.60.Fr, 11.10.Kk}

% ========================= BODY =========================
%\narrowtext

\section{Introduction}
\label{introduction}

According to the universality hypothesis, critical phenomena can be 
described in terms of quantities that do not depend on the microscopic details
of the  system, but only on global properties such as the dimensionality
and the symmetry of the order parameter, and the range of the interactions.
There exist several physical systems that are characterized by 
short-range interactions and 
an $N$-component order parameter. Because of universality, their
critical properties can be studied by using the Ginzburg-Landau 
$\phi^4$ Hamiltonian and by employing standard
field-theoretic renormalization-group techniques.
When the order parameter has only one component, one obtains the Ising
universality class that describes for instance 
the liquid-vapor transition in simple 
fluids and the transitions of multicomponent fluid systems; in this case
the density plays the role of the order parameter.
The two-component model (XY model) describes 
the helium superfluid transition, the Meissner transition in type-II
superconductors  and some transitions in liquid crystals, while the 
limit $N\to0$ gives 
the infinite-length properties of dilute polymers in a good solvent. 

The critical properties of many magnetic materials are also 
computed using the $N$-component Ginzburg-Landau Hamiltonian.
Uniaxial (anti-)ferromagnets should be described by the Ising universality
class ($N=1$), while magnets with easy-plane anisotropy should belong to
the XY universality class. Ferromagnets with cubic symmetry are often
described in terms of the $N=3$ Hamiltonian. However, this is correct 
if the non-rotationally invariant interactions that have only the 
reduced symmetry of the lattice are irrelevant in the renormalization-group
sense. Standard considerations
based on the canonical dimensions of the operators indicate that 
there are two terms that one may add to the Hamiltonian and that are 
cubic invariant: a cubic hopping term $\sum_{\mu=1,3} (\partial_\mu\phi_\mu)^2$
and a cubic interaction term $\sum_{\mu=1,3} \phi_\mu^4$.
The first interaction was studied in 
Refs. \cite{Aharony-73-3,Bruce-74,N-T-75,Nattermann-76,Aharony-76}.
A two-loop $O(\epsilon^2)$ calculation indicates that it is irrelevant 
at the symmetric point, although it induces slowly-decaying
crossover effects.
We will not consider it here 
--- although it would be worthwhile to perform a more systematic study ---
since the second term already introduces significant changes in the 
critical behavior of the system.
We will therefore consider a three-dimensional $\phi^4$ theory with two quartic couplings
\cite{Aharony-73,Aharony-76}:
\begin{equation}
{\cal H} = \int d^d x 
\left\{ {1\over 2} \sum_{i=1}^{N}
      \left[ (\partial_\mu \phi_i)^2 +  r \phi_i^2 \right] + 
 {1\over 4!} \sum_{i,j=1}^N \left( u_0 + v_0 \delta_{ij} \right)
\phi^2_i \phi^2_j \right\} 
\label{Hphi4}
\end{equation}
The added cubic term breaks explicitly the O($N$) invariance of the model,
leaving a residual discrete cubic symmetry given by the reflections
and permutations of the field components.  

The model described by the Hamiltonian (\ref{Hphi4}) 
has been extensively studied. It
has four fixed points~\cite{Aharony-73,Aharony-76}:
the trivial Gaussian one, the Ising one in which the $N$ components of the 
field decouple,
the O($N$)-symmetric and the  cubic fixed points.

The Gaussian fixed point is always unstable, and so is 
the Ising fixed point~\cite{Sak-74}.
Indeed, in the latter case, it is natural to interpret Eq. (\ref{Hphi4}) 
as the Hamiltonian of $N$ Ising-like systems coupled by the 
$O(N)$-symmetric term. But this interaction is the sum of the products 
of the energy operators of the different Ising systems.
Therefore, at the Ising fixed point, the crossover exponent 
associated to the O($N$)-symmetric quartic term should be given by
the specific-heat critical exponent $\alpha_I$ of the Ising model,
independently of $N$. 
Since $\alpha_I$ is positive, indeed
$\alpha_I = 0.1099(7)$
(see e.g. Ref.~\cite{C-P-R-V-99-1} and references therein), 
the Ising fixed point is unstable.

\begin{figure}[tb]
\vspace{-2cm}
\centerline{\psfig{width=15truecm,angle=-90,file=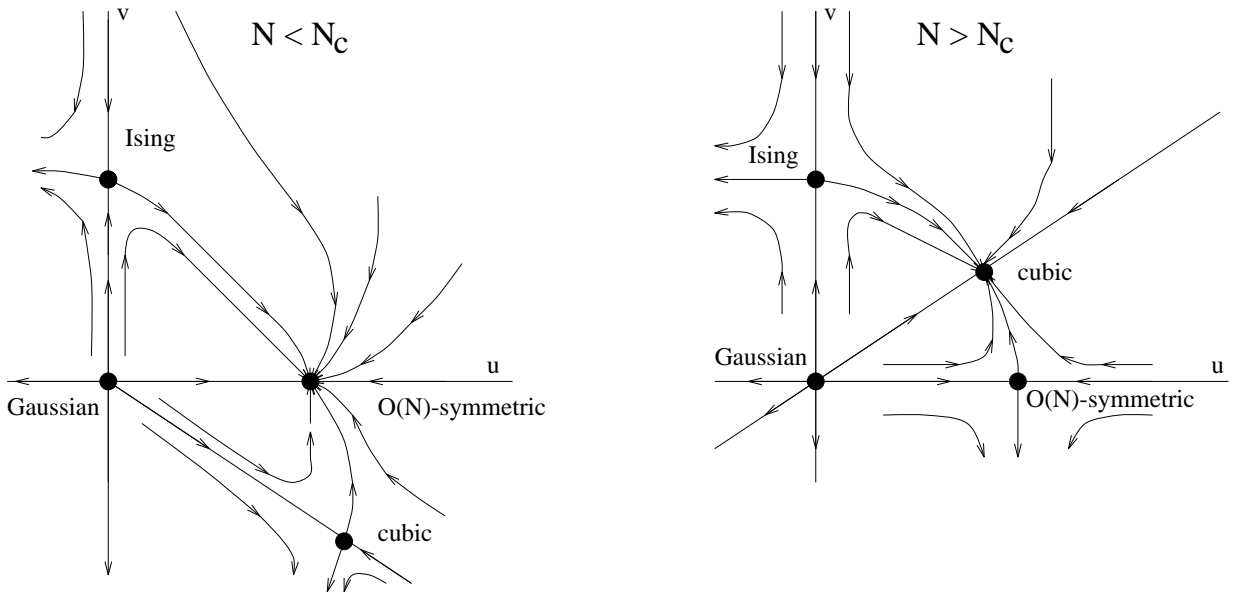}}
\vspace{-2cm}
\caption{
Renormalization-group flow in the coupling plane $(u,v)$ for
$N<N_c$ and $N>N_c$.
}
\label{rgflow}
\end{figure}

While the  Gaussian and the Ising fixed points are unstable
for any number of components $N$,
the stability properties
of the O($N$)-symmetric and of the cubic fixed points depend on $N$.
For sufficiently small values of $N$, $N<N_c$, the 
O($N$)-symmetric fixed point is stable and the cubic one is unstable.
For $N>N_c$, the opposite is true:
the renormalization-group flow is driven towards the cubic fixed point,
which now describes the generic critical behavior of the system.
The O($N$)-symmetric point corresponds to a tricritical transition.
Figure~\ref{rgflow} sketches the 
flow diagram in the two cases $N<N_c$ and $N>N_c$.
At $N=N_c$, the two fixed points should coincide, and
logarithmic corrections to the O($N$)-symmetric critical exponents are expected.
Outside the attraction domain of the fixed points, the flow goes away
towards more negative values of $u$ and/or $v$ and finally
reaches the region where the quartic interaction no longer 
satisfies the stability
condition. These trajectories should be related to
first-order phase transitions \cite{footnote1,footnote2}.

If $N>N_c$, the cubic anisotropy is relevant and therefore the critical
behavior of the system is not described by the Heisenberg isotropic 
Hamiltonian. If the cubic interaction favors the alignment of 
the spins along the diagonals of the cube,
i.e. for a positive coupling $v_0$,
the critical behavior is controlled by the cubic
fixed point and the cubic symmetry is retained even at the 
critical point. On the other hand, if the system tends to 
magnetize along the cubic axes --- this corresponds to a negative
coupling $v_0$ --- then the system undergoes a first-order phase 
transition \cite{Wallace-73,Aharony-76,Aharony_77,S-D-99}.
Moreover, since the symmetry is discrete, 
there are no Goldstone excitations in the low-temperature phase.
The longitudinal and the transverse
susceptibilities are finite for $T<T_c$ and $H\rightarrow 0$,
and diverge as $|t|^{-\gamma}$ for $t\propto T - T_c \rightarrow 0$
\cite{footnote3}.

In the limit $N\to\infty$, keeping $Nu$ and $v$ fixed, one can derive 
exact expressions for the exponents at the cubic fixed point.
Indeed, in this limit the model can be reinterpreted as 
a constrained Ising model~\cite{Emery-75}, leading to a 
Fisher renormalization of the Ising critical exponents~\cite{Fisher-68}.
One has~\cite{Aharony-73-2,Emery-75,Aharony-76}: 
\begin{equation}
\eta = \eta_I+O\left( {1\over N}\right),
\qquad\qquad   \nu = {\nu_I\over 1-\alpha_I}+O\left( {1\over N}\right),
\label{largen}
\end{equation}
where $\eta_I$, $\nu_I$, and $\alpha_I$ are the critical exponents of the
Ising model (see e.g. Ref.~\cite{C-P-R-V-99-1} and references therein
for recent estimates of the
Ising critical exponents).

If $N < N_c$, the cubic term in the Hamiltonian is irrelevant, and therefore, 
it generates 
only scaling corrections $|t|^{\Delta_c}$ with
$\Delta_c>0$. However,
their presence leads to important physical consequences. 
For instance,
the transverse susceptibility at the coexistence curve (i.e.
for $T<T_c$ and $H\rightarrow 0$), which is divergent in the 
O($N$)-symmetric case, is now finite and
diverges only at $T_c$ as $|t|^{-\gamma-\Delta_c}$ 
\cite{Wallace-73,K-W-73,B-A-75,Aharony-76,B-L-Z-76}. 
In other words, below $T_c$,
the cubic term is a ``dangerous'' irrelevant operator.
Note that for $N$ sufficiently close to $N_c$,
irrespective of which fixed point is the stable one,
the irrelevant interaction
bringing from the unstable to the stable fixed point gives rise to  very
slowly decaying corrections to the leading scaling behavior. 

In three dimensions, a simple argument 
based on the symmetry of the two-component cubic model~\cite{Korz-76}
shows that the cubic fixed point is unstable for $N=2$, so that  $N_c>2$.
Indeed, for $N=2$, a $\pi/4$ internal rotation, i.e.
\begin{equation}
(\phi_1,\phi_2) 
\longrightarrow  {1\over \sqrt{2}} (\phi_1+\phi_2,\phi_1-\phi_2),
\label{sym1}
\end{equation}
maps the cubic Hamiltonian (\ref{Hphi4}) into a new one of the same form 
but with new couplings $(u_0',v_0')$ given by
\begin{equation}
u_0' = u_0+\case{3}{2}v_0, \qquad \qquad v_0' = -v_0.
\label{sym2}
\end{equation}
This symmetry maps the Ising fixed point onto the cubic one. Therefore,
the two fixed points describe the same theory and have the same stability.
Since the Ising point is unstable, the cubic point is unstable too, 
so that the stable point is the isotropic one.
In two dimensions, this is no longer true. Indeed, one expects 
the cubic interaction to be truly marginal for $N=2$~\cite{J-K-K-N-77,N-R-82} and relevant for 
$N>2$ \cite{footnote4}, so that $N_c = 2$ in two dimensions.

During the years, the model (\ref{Hphi4}) has been the object of several studies
\cite{Aharony-73,K-W-73,G-K-W-72,N-K-F-74,B-L-Z-74,N-T-75,Nattermann-76,Y-H-77,D-R-79,F-V-C-81,N-R-82,M-S-87,%
Shpot-89,M-S-S-89,K-S-95,K-T-95,K-T-S-97,S-A-S-97,C-H-98,M-V-98,Varnashev-99}. 
In the 70's several computations were done using the 
$\epsilon$-expansion \cite{Aharony-73,K-W-73,N-K-F-74,B-L-Z-74};
they predicted $3< N_c < 4$, indicating that cubic ferromagnets
are described by the $O(N)$-invariant Heisenberg model. 
However, recent studies have questioned these conclusions.
Field-theoretic studies,
based on the analysis of  the three-loop \cite{M-S-87,Shpot-89} and
four-loop series~\cite{M-S-S-89,Varnashev-99}
in fixed  dimension,
and of the five-loop expansion in powers of $\epsilon=4-d$ 
\cite{K-S-95,K-T-95,K-T-S-97,S-A-S-97,Varnashev-99} 
suggest that $N_c\lesssim 3$, although they do not seem to be 
conclusive in excluding the value $N_c=3$.
On the other hand, the results of  Ref.~\cite{C-H-98},    
obtained from Monte Carlo simulations using finite-size scaling 
techniques, are perfectly consistent with the value $N_c\approx3$. 
The authors of Ref.~\cite{C-H-98} even suggest that $N_c=3$ exactly.

A further study of this issue is therefore of particular relevance 
for the ferromagnetic
materials characterized by an order parameter with $N=3$. For this purpose 
we extended the perturbative expansions of the $\beta$-functions and 
of the exponents to six loops
in the framework of the fixed-dimension field-theoretic approach~\cite{Parisi-80}.
These  perturbative expansions are only asymptotic. Nonetheless, 
accurate results can be obtained by employing resummation techniques that 
use their Borel summability \cite{bores} and the knowledge
of the large-order behavior \cite{Lipatov-77,B-L-Z-77}.
For this reason, we have also computed the singularity of the Borel transform
that is closest to the origin, extending the calculations of Refs. 
\cite{Lipatov-77,B-L-Z-77}.

\begin{table}[tbp]
%\squeezetable
\caption{
Summary of the results in the literature. 
The values of the exponents refer to $N=3$.
The subscripts ``$s$" and ``$c$" indicate that the exponent 
is related to the symmetric and to the cubic fixed point respectively.
}
\label{literature}
\begin{tabular}{ccc}
\multicolumn{1}{c}{}&
\multicolumn{1}{c}{method}&
\multicolumn{1}{c}{results}\\
\tableline \hline
Ref.\cite{N-K-F-74}, 1974 & 
      $\epsilon$-exp.: $O(\epsilon^3)$  & $N_c\simeq 3.128$  \\
Ref.\cite{Y-H-77}, 1977 & approx. RG  & $\nu_s \,\omega_{2,s}=-0.11,\;\;
     N_c \simeq 2.3 $  \\
Ref.\cite{F-V-C-81}, 1981 & H.T. exp.: $O(\beta^{10})$ & $\nu_s \,\omega_{2,s}
= -0.63(10),\;\;N_c<3 $  \\
Ref.\cite{N-R-82}, 1982 & scaling-field & $N_c\simeq 3.38$    \\
Ref.\cite{M-S-S-89}, 1989 & $d=3$ exp.: $O(g^4)$  
   & $\omega_{2,c}\simeq 0.008 ,\;\;N_c\simeq 2.91$  \\
Ref.\cite{K-S-95}, 1995 & 
  $\epsilon$-exp.: $O(\epsilon^5)$  &$N_c\simeq 2.958$ \\
Ref.\cite{K-T-S-97}, 1997 & $\epsilon$-exp.: $O(\epsilon^5)$ & $
   \omega_{2,s}=-0.00214,\;\;\omega_{2,c}=0.00213,\;\;N_c<3$ \\
Ref.\cite{S-A-S-97}, 1997 & 
   $\epsilon$-exp.: $O(\epsilon^5)$ &$N_c\simeq 2.86$ \\
Ref.\cite{C-H-98}, 1998 & 
     Monte Carlo & $\omega_{2,s}=0.0007(29),\;\;N_c\approx 3$ \\
Ref.\cite{Varnashev-99}, 1999 & $d=3$ exp.: $O(g^4)$ &
    $\omega_{2,s}=-0.0081,\;\;\omega_{2,c}=0.0077,\;\;N_c=2.89(2)$\\
This work & $\epsilon$-exp.: $O(\epsilon^5)$ &
   $\omega_{2,s}=-0.003(4),\;\;\omega_{2,c}=0.006(4),\;\;N_c=2.87(5)$ \\
This work & $d=3$ exp.: $O(g^6)$ &
$\omega_{2,s}=-0.013(6),\;\;\omega_{2,c}=0.010(4),\;\;N_c=2.89(4)$
\end{tabular}
\end{table}

The analysis of the perturbative series has been done following closely
Ref. \cite{L-Z-77}. We have estimated errors 
using an algorithmic procedure, trying to be as conservative as possible.
This can be immediately realized by comparing our uncertainties with 
those previously quoted: even though our series are longer, the errors we 
report are sometimes larger than those of previous studies. 
Our results confirm previous field-theoretic studies: the $N=3$ isotropic
fixed point is indeed unstable and we estimate $N_c = 2.89(4)$ from the 
six-loop fixed-dimension expansion and $N_c = 2.87(5)$ from the 
reanalysis of the five-loop $\epsilon$-expansion. 
For comparison, in Table \ref{literature} we report our estimates 
together with previous determinations of $N_c$ and of the eigenvalues 
for $N=3$.
It should 
be noted that the estimates of the critical exponents do not essentially 
depend on which fixed point is the stable one. Moreover, the tiny difference
(smaller than the precision of our results, which is $\lesssim 1\%$ in the case of $\gamma$ and $\nu$) 
between the values at the two fixed points would be very difficult to 
observe, because of crossover effects decaying as $t^{\Delta}$ with
$\Delta= \omega_{2,c}\nu_c = 0.007(3)$. 
Large corrections to scaling appear also for $N=2$. Indeed,
at the XY fixed point (the stable one), we find
$\omega_2 = 0.103(8)$. Thus, even though the cubic interaction is 
irrelevant, it induces strong scaling corrections behaving 
as $t^\Delta$, $\Delta=\omega_2\nu\approx 0.06$. Therefore, crossover
effects are expected in this case, depending on the strength of the 
cubic term. Finally, we have checked the theoretical predictions for 
the model in the large-$N$ limit finding good agreement.

We want to mention that, in
the limit $N\to0$, the cubic model (\ref{Hphi4}) describes 
the Ising model with site-diluted disorder
\cite{Harris-Lubensky_74,Lubensky_75,Khmelnitskii_75}.
However, in this case, the perturbative expansion is not 
Borel summable \cite{Bray-etal_87,McKane_94,Alvarez-etal_99}. 
Therefore, it is not completely clear how to obtain meaningful results
from the perturbative series. An investigation of these problems will 
be presented elsewhere. 

The paper is organized as follows. In Sec. \ref{sec2} 
we present our calculation of the perturbative expansions to six loops in
$d=3$. We give the basic definitions, the six-loop series, 
and the singularity of the Borel transform. In Sec. \ref{sec3} 
we present the analysis of these expansions: we determine the stability of 
the fixed points and compute the exponents for several values of $N$. 
In Sec. \ref{sec4} we present a reanalysis of the $\epsilon$-expansion
five-loop series.  The new analysis differs from the previous ones 
in the fact that it uses the large-order behavior of the series at the 
cubic fixed point. Finally, in the Appendix we report a 
three-loop $\epsilon$-expansion
computation of the zero-momentum four-point couplings at the cubic fixed 
point in three dimensions.

\section{The fixed-dimension perturbative expansion in three dimensions}
\label{sec2}

\subsection{Renormalization of the $\phi^4$ theory with cubic anisotropy}
\label{sec2a}

The fixed-dimension $\phi^4$ field-theoretic approach~\cite{Parisi-80} 
provides an accurate description of the
critical properties of $O(N)$-symmetric models 
in the high-temperature phase (see e.g. Ref.~\cite{ZJ-book}).
The method can also be applied to the two-parameter 
cubic model~\cite{M-S-87}.  
The idea is to perform an expansion in powers of appropriately defined 
zero-momentum quartic couplings. 
In order to obtain estimates of the universal critical quantities, 
the perturbative series are resummed
exploiting their Borel summability, 
and then evaluated at the fixed-point values of the couplings. 

The theory is renormalized by introducing a set of zero-momentum conditions 
for the (one-particle irreducible) two-point 
and four-point correlation functions:
\begin{equation}
\Gamma^{(2)}_{ab}(p) = \delta_{ab} Z_\phi^{-1} \left[ m^2+p^2+O(p^4)\right],
\label{ren1}  
\end{equation}
\begin{equation}
\Gamma^{(4)}_{abcd}(0) = 
Z_\phi^{-2} m \left[  
{u\over 3}\left(\delta_{ab}\delta_{cd} + \delta_{ac}\delta_{bd} + 
                \delta_{ad}\delta_{bc} \right)
+ v \,\delta_{ab}\delta_{ac}\delta_{ad}\right].
\label{ren2}  
\end{equation}
They relate the second-moment mass $m$, and the zero-momentum
quartic couplings $u$ and $v$ to the corresponding Hamiltonian parameters
$r$, $u_0$ and $v_0$:
\begin{equation}
u_0 = m u Z_u Z_\phi^{-2},\qquad\qquad
v_0 = m v Z_v Z_\phi^{-2}.
\end{equation}
In addition, one introduces the function $Z_t$ that is defined by the relation
\begin{equation}
\Gamma^{(1,2)}_{ab}(0) = \delta_{ab} Z_t^{-1},
\label{ren3}
\end{equation}
where $\Gamma^{(1,2)}$ is the (one-particle irreducible)
two-point function with an insertion of $\case{1}{2}\phi^2$.

From the pertubative expansion of the correlation functions
$\Gamma^{(2)}$, $\Gamma^{(4)}$ and $\Gamma^{(1,2)}$ and 
the above relations, one derives the functions $Z_\phi(u,v)$, 
$Z_u(u,v)$, $Z_v(u,v)$, $Z_t(u,v)$ as a double expansion in $u$ and $v$.

The fixed points of the theory are given by 
the common  zeros of the $\beta$-functions
\begin{eqnarray}
\beta_u(u,v) &=& m \left. {\partial u\over \partial m}\right|_{u_0,v_0} ,
\nonumber \\
\beta_v(u,v) &=& m \left. {\partial v\over \partial m}\right|_{u_0,v_0} .
\end{eqnarray}
The stability properties of the fixed points are controlled  by the 
eigenvalues $\omega_i$ of the matrix 
\begin{equation}
\Omega =\left(\matrix{\frac{\partial \beta_u(u,v)}{\partial u}
 & \frac{\partial \beta_u(u,v)}{\partial v}
 \cr\frac{\partial \beta_v(u,v)}{\partial u}
&  \frac{\partial \beta_v(u,v)}{\partial v}}\right)\; ,
\end{equation}
computed at the given fixed point:
a fixed point is stable if both eigenvalues are positive.
The eigenvalues $\omega_i$ are related to
the leading scaling corrections, which vanish as
$\xi^{-\omega_i}\sim |t|^{\Delta_i}$ where $\Delta_i=\nu\omega_i$.

One also introduces the functions
\begin{eqnarray}
\eta_\phi(u,v) &=& \left. {\partial \ln Z_\phi \over \partial \ln m}
         \right|_{u_0,v_0}
= \beta_u {\partial \ln Z_\phi \over \partial u} +
\beta_v {\partial \ln Z_\phi \over \partial v} ,\\
\eta_t(u,v) &=& \left. {\partial \ln Z_t \over \partial \ln m}
         \right|_{u_0,v_0}
= \beta_u {\partial \ln Z_t \over \partial u} +
\beta_v {\partial \ln Z_t \over \partial v}.
\end{eqnarray}
Finally, the critical exponents are obtained from
\begin{eqnarray}
\eta &=& \eta_\phi(u^*,v^*),
\label{eta_fromtheseries} \\
\nu &=& \left[ 2 - \eta_\phi(u^*,v^*) + \eta_t(u^*,v^*)\right] ^{-1},
\label{nu_fromtheseries} \\
\gamma &=& \nu (2 - \eta).
\label{gamma_fromtheseries} 
\end{eqnarray}

\subsection{The six-loop perturbative series}
\label{sec2b}

We have computed the perturbative expansion of the 
correlation functions (\ref{ren1}), (\ref{ren2}) and (\ref{ren3})
to six loops. The diagrams contributing to the two-point and 
four-point functions to six-loop order
are reported in Ref.~\cite{N-M-B-77}: they are approximately one thousand, 
and it is therefore necessary to handle them with a symbolic manipulation
program. For this purpose, we wrote a package in 
{\sc Mathematica} \cite{Wolfram}.
It generates the diagrams using the algorithm described in Ref.~\cite{Heap-66},
and computes the symmetry and group factors of 
each of them. We did not calculate the integrals associated to each diagram,
but we used the numerical results compiled in Ref.~\cite{N-M-B-77}.
Summing all contributions we determined the renormalization constants
and all renormalization-group functions.

We report our results in terms of the rescaled couplings~\cite{B-N-G-M-77}
\begin{equation}
u \equiv  {16 \pi\over 3} \; R_N \; \bar{u},\qquad\qquad
v \equiv   {16 \pi\over 3} \; \bar{v} ,
\label{resc}
\end{equation}
where $R_N = 9/(8+N)$,
so that the $\beta$-functions associated to $\bar{u}$ and $\bar{v}$ have the 
form $\beta_{\bar{u}}(\bar{u},0) = -\bar{u} + \bar{u}^2 + O(\bar{u}^3)$ and 
$\beta_{\bar{v}}(0,\bar{v}) = -\bar{v} + \bar{v}^2 + O(\bar{v}^3)$. 

The resulting series are:
\begin{eqnarray}
\beta_{\bar{u}}  =&& -\bar{u} + \bar{u}^2 + {2\over 3} \bar{u} \bar{v} 
- {4(190+41N)\over 27(8+N)^2} \bar{u}^3 \label{bu}\\
&&- {400\over 81(8+N)} \bar{u}^2 \bar{v} - {92\over 729}\bar{u}\bar{v}^2 +
\bar{u} \sum_{i+j\geq 3} b^{(u)}_{ij} \bar{u}^i \bar{v}^j, \nonumber 
\end{eqnarray}
\begin{eqnarray}
\beta_{\bar{v}}  = && -\bar{v} + \bar{v}^2 + {12\over 8+N} \bar{u} \bar{v} - 
{308\over 729} \bar{v}^3 
- {832\over 81(8+N)} \bar{u}\bar{v}^2 \label{bv} \\
&&- {4(370+23N)\over 27(8+N)^2} \bar{u}^2 \bar{v} +
\bar{v} \sum_{i+j\geq 3} b^{(v)}_{ij} \bar{u}^i \bar{v}^j, \nonumber 
\end{eqnarray}
\begin{equation}
\eta_\phi = 
{8 (2 + N)\over 27(8+N)^2 } \bar{u}^2 + {16\over 81(8+N) } \bar{u} \bar{v} + 
{8 \over 729 } \bar{v}^2 + \sum_{i+j\geq 3} e^{(\phi)}_{ij} \bar{u}^i \bar{v}^j,
\label{etaphi}
\end{equation}
\begin{equation}
\eta_t = -{ (2+N)\over (8+N) } \bar{u} -{1\over 3} \bar{v} +
{ 2(2+N)\over (8+N)^2 } \bar{u}^2 + 
{ 4 \over 3 (8+N) } \bar{u} \bar{v} 
+ {2\over 27} \bar{v}^2 +
\sum_{i+j\geq 3} e^{(t)}_{ij} \bar{u}^i \bar{v}^j.
\label{etat}
\end{equation}
For $3\leq i+j\leq  6$,
the coefficients $b^{(u)}_{ij}$, $b^{(v)}_{ij}$,
$e^{(\phi)}_{ij}$ and $e^{(t)}_{ij}$ 
are reported in Tables~\ref{betauc}, \ref{betavc},
\ref{ephi} and \ref{et} respectively.

We have performed several checks of our calculations:

(i) $\beta_{\bar{u}}(\bar{u},0)$, $\eta_\phi(\bar{u},0)$ 
and $\eta_t(\bar{u},0)$ reproduce
the corresponding functions of the O($N$)-symmetric
model~\cite{B-N-G-M-77,A-S-95};

(ii) $\beta_{\bar{v}}(0,\bar{v})$, $\eta_\phi(0,\bar{v})$ and 
$\eta_t(0,\bar{v})$ reproduce
the corresponding functions of the Ising-like ($N=1$) $\phi^4$ theory;

(iii) The following relations hold for $N=1$: 
\begin{eqnarray}
&& \beta_{\bar{u}}(u,x-u) + \beta_{\bar{v}}(u,x-u) =\beta_{\bar{v}}(0,x),\\
&&\eta_\phi(u,x-u) = \eta_\phi(0,x),\nonumber\\
&&\eta_t(u,x-u) = \eta_t(0,x).\nonumber
\end{eqnarray}

(iv) For $N=2$, using the symmetry 
(\ref{sym1}) and (\ref{sym2}), and 
taking into account the rescalings (\ref{resc}), one can easily obtain
the identities 
\begin{eqnarray}
&& \beta_{\bar{u}}(\bar{u}+\case{5}{3}\bar{v},-\bar{v}) + 
{5\over 3}\beta_{\bar{v}}(\bar{u}+\case{5}{3}\bar{v},-\bar{v}) = 
\beta_{\bar{u}}(\bar{u},\bar{v}),\\
&& \beta_{\bar{v}}(\bar{u}+\case{5}{3}\bar{v},-\bar{v}) = -\beta_{\bar{v}}(\bar{u},\bar{v}), \nonumber\\
&&\eta_\phi(\bar{u}+\case{5}{3}\bar{v},-\bar{v}) = \eta_\phi(\bar{u},\bar{v}),\nonumber\\
&&\eta_t(\bar{u}+\case{5}{3}\bar{v},-\bar{v}) = \eta_t(\bar{u},\bar{v}).\nonumber
\end{eqnarray}
These relations are exactly satisfied by our six-loop series.
Note that, since the Ising fixed point is $(0,g^*_I)$ with
$g^*_I=1.402(2)$~\cite{C-P-R-V-99-1}, the above symmetry gives us 
the location of the cubic fixed point: $(\case{5}{3} g^*_I,-g^*_I)$.

(v) In the large-$N$ limit 
the critical exponents of the cubic fixed point 
are related to those of the Ising model:
$\eta=\eta_I$ and $\nu = \nu_I/(1-\alpha_I)$.
One can easily see
that, for $N\to\infty$,
$\eta_\phi(u,v) = \eta_I(v)$, where $\eta_I(v)$ is 
the perturbative series that determines the exponent $\eta$ of the Ising model.
Therefore, the first relation is trivially true. On the other hand,
the second relation $\nu = \nu_I/(1-\alpha_I)$ is not identically
satisfied by the series, and is verified only at the critical point
\cite{footnote5}.

We finally note that our series are in agreement with the five-loop results
appeared recently in Ref.~\cite{P-S-00}.

\begin{table}[tbp]
\squeezetable
\caption{
The coefficients $b^{(u)}_{ij}$, cf. Eq. (\ref{bu}).
}
\label{betauc}
\begin{tabular}{cl}
\multicolumn{1}{c}{$i,j$}&
\multicolumn{1}{c}{$R_N^{-i} b^{(u)}_{ij}$}\\
\tableline \hline
3,0 &$ 0.27385517 + 0.075364029\,N + 0.0018504016\,{N^2}$\\
2,1 &$ 0.67742325 + 0.027353409\,N$\\
1,2 &$ 0.4154565 + 0.0025592148\,N$\\
0,3 &$ 0.090448951$\\\hline
4,0 &$ -0.27925724 - 0.091833749\,N - 0.0054595646\,{N^2} + 
     0.000023722893\,{N^3}$\\
3,1 &$ -0.94383662 - 0.083252807\,N + 0.00061860174\,{N^2}$\\
2,2 &$ -0.96497888 - 0.012460145\,N$\\
1,3 &$ -0.42331874 - 0.0017709429\,N$\\
0,4 &$ -0.075446692$\\\hline
5,0 &$ 0.35174477 + 0.13242502\,N + 0.011322026\,{N^2} + 
     0.000054833719\,{N^3} + 8.6768933\,{{10}^{-7}}\,{N^4}$\\
4,1 &$ 1.5209008 + 0.19450536\,N + 0.0011078614\,{N^2} + 0.000031779782\,{N^3}   $\\
3,2 &$ 2.2073347 + 0.065336326\,N + 0.0003564925\,{N^2}$\\
2,3 &$ 1.5315693 + 0.010676901\,N$\\
1,4 &$ 0.56035196 + 0.0013469481\,N$\\
0,5 &$ 0.087493302$\\\hline
6,0 &$ -0.51049889 - 0.21485252\,N - 0.023839375\,{N^2} - 
    0.00050021682\,{N^3} + 2.0167763\,{{10}^{-6}}\,{N^4} + 
     4.4076733\,{{10}^{-8}}\,{N^5}$\\
5,1 &$ -2.6984083 - 0.45068252\,N - 0.010821468\,{N^2} + 
     0.00005796668\,{N^3} + 2.0515456\,{{10}^{-6}}\,{N^4}$\\
4,2 &$ -5.1135549 - 0.26769177\,N - 0.0006311751\,{N^2} + 
     0.000019413374\,{N^3}$\\
3,3 &$ -4.9317312 - 0.067574712\,N + 0.000028278087\,{N^2}$\\
2,4 &$ -2.754683 - 0.0095836704\,N$\\
1,5 &$ -0.86229463 - 0.001856332\,N$\\
0,6 &$ -0.1179508$
\end{tabular}
\end{table}

\begin{table}[tbp]
\squeezetable
\caption{The coefficients $b^{(v)}_{ij}$, cf. Eq. (\ref{bv}).
}
\label{betavc}
\begin{tabular}{cl}
\multicolumn{1}{c}{$i,j$}&
\multicolumn{1}{c}{$R_N^{-i} b^{(v)}_{ij}$}\\
\tableline \hline
3,0 &$ 0.64380517 + 0.05741276\,N - 0.0017161966\,{N^2}$\\
2,1 &$ 1.6853305 + 0.0030714114\,N$\\
1,2 &$ 1.3138294$\\
0,3 &$ 0.3510696$\\\hline
4,0 &$ -0.76706177 - 0.089054667\,N + 0.000040711369\,{N^2} - 
     0.000087586118\,{N^3}$\\
3,1 &$ -2.7385841 - 0.049218875\,N - 0.00002623469\,{N^2}$\\
2,2 &$ -3.3477204 + 0.0075418394\,N$\\
1,3 &$ -1.8071874$\\
0,4 &$ -0.37652683$\\\hline
5,0 &$ 1.0965348 + 0.15791293\,N + 0.0023584631\,{N^2} - 
     0.000061471346\,{N^3} - 5.3871247\,{{10}^{-6}}\,{N^4}$\\
4,1 &$ 4.9865485 + 0.17572792\,N - 0.0020718369\,{N^2} - 0.000019382912\,{N^3}   $\\
3,2 &$ 8.3645284 + 0.0039620562\,N + 0.00021363122\,{N^2}$\\
2,3 &$ 6.8946012 - 0.0230874\,N$\\
1,4 &$ 2.8857918$\\
0,5 &$ 0.49554751$\\\hline
6,0 &$ -1.7745533 - 0.30404316\,N - 0.0094338079\,{N^2} + 
     0.000066993864\,{N^3} - 6.5724895\,{{10}^{-6}}\,{N^4} - 
     3.753114\,{{10}^{-7}}\,{N^5}$\\
5,1 &$ -9.8298296 - 0.53384955\,N + 0.0022033252\,{N^2} - 
     0.00013066822\,{N^3} - 2.5959429\,{{10}^{-6}}\,{N^4}$\\
4,2 &$ -21.073538 - 0.16628697\,N - 0.000014827682\,{N^2} + 
     4.4988524\,{{10}^{-6}}\,{N^3}$\\
3,3 &$ -23.569724 + 0.095716867\,N - 0.00083903999\,{N^2}$\\
2,4 &$ -14.927998 + 0.0486813\,N$\\
1,5 &$ -5.1298717$\\
0,6 &$ -0.74968893$
\end{tabular}
\end{table}

\begin{table}[tbp]
\squeezetable
\caption{
The coefficients $e^{(\phi)}_{ij}$, cf. Eq.(\ref{etaphi}).
}
\label{ephi}
\begin{tabular}{cl}
\multicolumn{1}{c}{$i,j$}&
\multicolumn{1}{c}{$R_N^{-i} e^{(\phi)}_{ij}$}\\
\tableline \hline
3,0 &$ 0.00054176134 + 0.00033860084\,N + 0.000033860084\,{N^2}$\\
2,1 &$ 0.002437926 + 0.00030474076\,N$\\
1,2 &$ 0.0027426668$\\
0,3 &$ 0.00091422227$\\\hline
4,0 &$ 0.00099254838 + 0.00070251807\,N + 0.0001018116\,{N^2} - 
     6.5516886\,{{10}^{-7}}\,{N^3}$\\
3,1 &$ 0.0059552903 + 0.0012374633\,N - 7.8620264\,{{10}^{-6}}\,{N^2}$\\
2,2 &$ 0.01046567 + 0.00031166693\,N$\\
1,3 &$ 0.0071848915$\\
0,4 &$ 0.0017962229$\\\hline
5,0 &$ -0.00036659735 - 0.0002572117\,N - 0.000032026611\,{N^2} + 
     2.2430702\,{{10}^{-6}}\,{N^3} - 1.1094045\,{{10}^{-7}}\,{N^4}$\\
4,1 &$ -0.0027494801 - 0.00055434769\,N + 0.000036974267\,{N^2} - 
     1.6641067\,{{10}^{-6}}\,{N^3}$\\
3,2 &$ -0.0064696069 - 0.000070210701\,N + 2.7823882\,{{10}^{-6}}\,{N^2}$\\
2,3 &$ -0.0066046286 + 0.00006759333\,N$\\
1,4 &$ -0.0032685176$\\
0,5 &$ -0.00065370353$\\\hline
6,0 &$ 0.00069568037 + 0.00056585941\,N + 0.00012057302\,{N^2} + 
     5.7466979\,{{10}^{-6}}\,{N^3} - 3.8385183\,{{10}^{-8}}\,{N^4} - 
     1.0441273\,{{10}^{-8}}\,{N^5}$\\
5,1 &$ 0.0062611234 + 0.001962173\,N + 0.00010407066\,{N^2} - 
     3.1504745\,{{10}^{-7}}\,{N^3} - 1.8794292\,{{10}^{-7}}\,{N^4}$\\
4,2 &$ 0.018957129 + 0.0018483045\,N + 0.000012482494\,{N^2} - 
     7.5537784\,{{10}^{-7}}\,{N^3}$\\
3,3 &$ 0.027158805 + 0.00059570458\,N + 1.7043408\,{{10}^{-6}}\,{N^2}$\\
2,4 &$ 0.020775681 + 0.000041479576\,N$\\
1,5 &$ 0.0083268641$\\
0,6 &$ 0.0013878107$
\end{tabular}
\end{table}

\begin{table}[tbp]
\squeezetable
\caption{
The coefficients $e^{(t)}_{ij}$, cf. Eq. (\ref{etat}).
}
\label{et}
\begin{tabular}{cl}
\multicolumn{1}{c}{$i,j$}&
\multicolumn{1}{c}{$R_N^{-i} e^{(t)}_{ij}$}\\
\tableline \hline
3,0 &$ -0.025120499 - 0.016979919\,N - 0.0022098349\,{N^2}$\\
2,1 &$ -0.11304225 - 0.019888514\,N$\\
1,2 &$ -0.13037154 - 0.0025592148\,N$\\
0,3 &$ -0.044310253$\\\hline
4,0 &$ 0.021460047 + 0.015690833\,N + 0.0024059273\,{N^2} - 
     0.000037238563\,{N^3}$\\
3,1 &$ 0.12876028 + 0.029764853\,N - 0.00044686275\,{N^2}$\\
2,2 &$ 0.22779178 + 0.0093256378\,N$\\
1,3 &$ 0.15630733 + 0.0017709429\,N$\\
0,4 &$ 0.039519569$\\\hline
5,0 &$ -0.022694287 - 0.017985168\,N - 0.0035835384\,{N^2} - 
     0.00013566164\,{N^3} - 1.699309\,{{10}^{-6}}\,{N^4}$\\
4,1 &$ -0.17020715 - 0.049785186\,N - 0.0019839454\,{N^2} - 
     0.000025489635\,{N^3}$\\
3,2 &$ -0.40917573 - 0.034538379\,N - 0.00028943185\,{N^2}$\\
2,3 &$ -0.43464785 - 0.0093557007\,N$\\
1,4 &$ -0.22065482 - 0.0013469481\,N$\\
0,5 &$ -0.044400355$\\\hline
6,0 &$ 0.029450619 + 0.024874579\,N + 0.005728397\,{N^2} + 
     0.00031557863\,{N^3} - 5.858689\,{{10}^{-6}}\,{N^4} - 
     1.0373506\,{{10}^{-7}}\,{N^5}$\\
5,1 &$ 0.26505557 + 0.091343427\,N + 0.0058838593\,{N^2} - 
     0.00010172194\,{N^3} - 1.8672312\,{{10}^{-6}}\,{N^4}$\\
4,2 &$ 0.80694662 + 0.09798244\,N + 0.00053192615\,{N^2} - 
     0.000012819748\,{N^3}$\\
3,3 &$ 1.1638111 + 0.043471034\,N - 0.000017867101\,{N^2}$\\
2,4 &$ 0.89481393 + 0.010634231\,N$\\
1,5 &$ 0.36032293 + 0.001856332\,N$\\
0,6 &$ 0.060363211$
\end{tabular}
\end{table}

\subsection{Singularity of the Borel transform}
\label{sec2c}
 
Since field-theoretic  perturbative expansions are asymptotic, 
the resummation of the series is essential
to obtain accurate estimates of the physical quantities.
In three-dimensional $\phi^4$ theories one exploits their Borel summability
\cite{bores} and the knowledge of the large-order behavior of the expansion 
(see e.g.  Ref.~\cite{ZJ-book}).

In the case of the O($N$)-symmetric $\phi^4$ theory, the expansion 
is performed in powers of the zero-momentum four-point coupling $g$.
The large-order behavior of the series $S(g) = \sum s_k g^k$
of any quantity is related to the singularity $g_b$ of the Borel transform
that is closest to the origin. Indeed, for large $k$,
\begin{equation}
s_k \sim k! \,(-a)^{k}\, k^b \,\left[ 1 + O(k^{-1})\right] \qquad\qquad
{\rm with}\qquad a = - 1/g_b.
\label{lobh}
\end{equation}
The value of $g_b$ depends only on the Hamiltonian,
while the exponent $b$ depends on which Green's function is considered.
If the perturbative expansion is Borel summable, then $g_b$ is negative.
The value of $g_b$ can be obtained from 
a steepest-descent calculation in which
the relevant saddle point is a finite-energy solution (instanton)
of the classical field equations with negative 
coupling~\cite{Lipatov-77,B-L-Z-77}.
If the Borel transform is singular for $g=g_b$, its expansion in powers 
of $g$ converges only for $|g|< |g_b|$.
An analytic extension 
can be obtained by a conformal mapping~\cite{L-Z-77}, such as
\begin{equation}
y(g) = {\sqrt{1 - g/g_b} - 1\over \sqrt{1 - g/g_b} + 1 }.
\end{equation}
In this way the Borel transform becomes a series in powers of $y(g)$
that converges for all positive values of $g$ provided that all 
singularities of the Borel transform are on the real negative 
axis~\cite{L-Z-77}.  For the O($N$)-symmetric theory accurate estimates 
(see e.g. Ref.~\cite{G-Z-98}) have been obtained resumming
the available series: the $\beta$-function~\cite{B-N-G-M-77} is known
up to six loops, while the functions $\eta_\phi$ and $\eta_t$ 
are known to seven loops~\cite{M-N-91}.
A subtle  point in this method is the
estimate of the uncertainty of the results.
Indeed the non-analyticity of the Callan-Symanzik $\beta$-function
at the fixed-point value $g^*$~\cite{Parisi-80,Nickel-82,Nickel-91,P-V-gr-98}
may cause a slow convergence of the estimates to the correct fixed-point value. 
This may lead to an underestimate of the
uncertainty which is usually derived from stability criteria.  
The reason is that
this resummation method approximates the $\beta$-function in the interval 
$[0,g^*]$ with a sum of analytic functions. Since, for $g=g^*$, the 
$\beta$-function is not analytic, the convergence at the endpoint
of the interval is slow. However, 
the comparison of these results with those obtained in other approaches 
shows that the above 
nonanalyticity causes only very small effects, 
which are negligible in most cases. 
See Refs.~\cite{P-V-gr-98,C-P-R-V-99-1} for a discussion of this issue.

In order to apply a resummation technique similar to 
that used in Ref.~\cite{L-Z-77}
to our six-loop series, i.e. in order to use a Borel transformation and 
a conformal mapping to get a convergent sequence of approximations,
we extended the large-order analysis to the cubic model.
In particular we considered the double expansion in $\bar{u}$ and $\bar{v}$ at 
fixed $z \equiv  \bar{v}/\bar{u}$. Then, we studied the large-order behavior 
of the resulting expansion in powers of $\bar{u}$.
This was done following the standard approach described 
for example in Refs.~\cite{B-L-Z-77,ZJ-book}, i.e. by studying 
the saddle-point solutions of the cubic model.
We will report the calculation elsewhere \cite{PelVic_inprep};
here we give only the results.

For $z \equiv \bar{v}/\bar{u}$ fixed and $N>1$, the singularity of the Borel transform 
closest to the origin, $\bar{u}_b$, is given by 
\begin{eqnarray}
{1\over \bar{u}_b} &= - a \left( R_N  + z \right)
\qquad\qquad & {\rm for} \qquad\qquad  0< z ,
\label{bsing} \\
{1\over \bar{u}_b} &= - a \left( R_N  + {1\over N} z \right)
\qquad\qquad & {\rm for} \qquad\qquad  0 > z > - {2 N R_N\over N+1},
\nonumber
\end{eqnarray}
where
\begin{equation}
a = 0.14777422...,\qquad\qquad R_{N} = {9\over N+8}.
\end{equation} 
Note that the series in powers of $\bar{u}$ keeping $z$ fixed
is not Borel summable for $\bar{u}>0$ and  $z< -R_N$.
This fact will not be a real limitation for us,
since we will only consider values of $z$ such that $\bar u_b<0$.
It should be noted that these results do not apply to the case $N=0$. 
Indeed, in this case, there are additional singularities in the 
Borel transform \cite{Bray-etal_87,McKane_94,Alvarez-etal_99}.

%One may also generalize the above calculation to different choices of the parameters,
%e.g., considering the series in terms of  $x=\bar{u}+c\bar{v}$, where $c$ is a constant,
%and keeping $z \equiv  \bar{v}/x$ fixed. In this case one finds 
%\begin{eqnarray}
%{1\over x_b} &= - a \left[ R_N  + z\left( {1\over N} - c R_N \right)\right]
%\qquad\qquad &{\rm for} \qquad\qquad -{2R_N N \over N+1} \left( 1 - 2c {R_N N\over N+1}\right)^{-1}< z <0,
%\label{bsingc} \\
%{1\over x_b} &= - a \left[ R_N + z\left( 1 - cR_N\right)\right]
%\qquad\qquad &{\rm for} \qquad\qquad z >0.
%\nonumber
%\end{eqnarray}

The exponent $b$ in Eq. (\ref{lobh}) is related to the number of 
symmetries broken by the classical solution~\cite{B-L-Z-77}.
It depends on the quantity considered. In the cubic model, for $v\not=0$, 
we have $b=5/2$ for the function $\eta_\phi$, and 
$b=7/2$ for the $\beta$-functions and $\eta_t$. 
For $v=0$, we recover the results of the O($N$)-symmetric model, that is 
$b=2 + N/2$ for $\eta_\phi$, and $b=3+N/2$ for 
the $\beta$-function and $\eta_t$~\cite{L-Z-77}.

\section{Analysis of the fixed-dimension six-loop series}
\label{sec3}

\begin{table}
%\squeezetable
\caption{
Estimates of the fixed-point value $\ub^*$ and of 
the critical exponent $\omega_2$ 
at the $O(N)$-symmetric fixed point for $N=3$.
The number $p$ indicates the number of loops included in the analysis,
the columns labelled by $\alpha$ and $b$ indicate the intervals 
of $\alpha$ and $b$ used, ``final" reports our
final estimate for the given analysis.}
\label{critical_exponents_symm_N3_gexp}
\begin{tabular}{lccllll}
& $\alpha$ &  $b$   &   $p=4$   &    $p=5$  &   $p=6$   & final \\
\hline
$\ub^*$& [0.5,2.5] & [5,11] & 1.400(6) & 1.392(4) & 1.392(1) & 1.392(2) \\
       & [0.5,2.5] & [5,13] & 1.400(6) & 1.392(4) & 1.392(1) & 1.392(2) \\
       & [0.5,2.5] & [3,13] & 1.406(17)& 1.389(7) & 1.394(4) & 1.394(8) \\
       & [$-$0.5,3.5] & [5,11] & 1.404(13)& 1.392(6) & 1.393(2) & 1.393(4) \\
%       & [$-$0.5,3.5] & [5,13] & 1.403(14)& 1.393(6) & 1.393(2) & 1.393(4) \\
\hline
$\omega_2$ & [0.5,2.5] & [5,11] & 
            $-$0.014(10) & $-$0.009(5) & $-$0.013(2) & $-$0.013(4) \\
           & [0.5,2.5] & [5,13] &
            $-$0.012(9)  & $-$0.009(4) & $-$0.012(2) & $-$0.012(4) \\
           & [0.5,2.5] & [3,13] &
            $-$0.017(22) & $-$0.007(12)& $-$0.014(7) & $-$0.014(14) \\
           & [$-$0.5,3.5] & [5,11] &
            $-$0.015(16) & $-$0.008(7) & $-$0.013(3) & $-$0.013(6) \\
%%           & [$-$0.5,3.5] & [5,13] &
%%          $-$0.013(16) & $-$0.009(7) & $-$0.013(3) & $-$0.013(6) \\
\end{tabular}
\end{table}

\begin{table}[tbp]
%\squeezetable
\caption{
Estimates of $\ub^*$ and of the eigenvalue $\omega_2$ at the $O(N)$-fixed point
for several values of $N$ and for different orders $p$ 
of the perturbative series. The last column reports the final estimate. }
\label{table_omega2_symmetricpoint}
\begin{tabular}{ccccccc}
$N$ & $\alpha$ & $b$ & $ p = 4$ & $p = 5$ & $p = 6$ & final \\
\hline
%% VECCHIE ANALISI 
%% $2$ &&&  0.101(3) &  0.104(3) &  0.103(1) &  0.103(2) \\
%% $3$ &&& $-$0.011(2) & $-$0.010(2) & $-$0.012(1) & $-$0.012(3) \\
%% $4$ &&& $-$0.107(2) & $-$0.108(2) & $-$0.111(1) & $-$0.111(4) \\
%% $8$ &&& $-$0.373(3) & $-$0.378(2) & $-$0.387(2) & $-$0.387(11) \\
%% $\infty$ &&& $-$1   &   $-$1      &   $-$1      &  $-$1\\
%% \hline
\multicolumn{7}{c}{$\overline{u}^*$} \\
\hline
2 & [$-$0.5,3.5] & [5,11] & 1.422(15) & 1.409(7) & 1.408(2) & 1.408(4) \\
3 & [$-$0.5,3.5] & [5,11] & 1.404(13) & 1.392(6) & 1.393(2) & 1.393(4) \\
4 & [0.5,4.5]    & [9,15] & 1.372(12) & 1.375(2) & 1.375(1) & 1.375(2) \\
8 & [0.5,4.5]    & [9,15] & 1.303(6)  & 1.304(1) & 1.305(1) & 1.305(2) \\
$\infty$ &       &        &           &          &          & 1        \\
\hline
\multicolumn{7}{c}{$\omega_2$} \\
\hline
2 & [$-$0.5,3.5] & [5,11] &
       0.099(20) & 0.107(11) & 0.103(4) & 0.103(8) \\
3 & [$-$0.5,3.5] & [5,11] &
    $-$0.015(16) & $-$0.008(7) & $-$0.013(3) & $-$0.013(6) \\
4 & [0.5,4.5]    & [9,15] &
    $-$0.105(10) & $-$0.109(3) & $-$0.111(2) & $-$0.111(4) \\
8 & [0.5,4.5]    & [9,15] &
    $-$0.371(8)  & $-$0.379(4) & $-$0.385(4) & $-$0.385(8) \\
$\infty$ &       &        &           &          &          &  $-$1 \\
\end{tabular}
\end{table}

In this Section we present the analysis of the six-loop series in fixed 
dimension. The resummation of the series has been done using the method of 
Ref. \cite{L-Z-77} and the expression for the Borel singularity we have given
in the previous Section. Explicitly, given a series
\begin{equation}
R(\ub,\vb) =\, \sum_{k=0} \sum_{h=0} R_{hk} \overline{u}^h \overline{v}^k,
\end{equation}
we have generated a new quantity $E({R})_p(\alpha,b;\ub,\vb)$
according to
\begin{equation}
E({R})_p(\alpha,b;\ub,\vb) = \sum_{k=0}^p 
  B_k(\alpha,b;\vb/\ub) \int_0^\infty dt\, t^b e^{-t} 
  {y(\ub t;\vb/\ub)^k\over [1 - y(\ub t;\vb/\ub)]^\alpha},
\end{equation}
where
\begin{equation}
y(x;z) = {\sqrt{1 - x/\overline{u}_b(z)} - 1\over 
          \sqrt{1 - x/\overline{u}_b(z)} + 1},
\end{equation}
and $\overline{u}_b(z)$ is defined in Eq. (\ref{bsing}). The coefficients 
$B_k(\alpha,b;\vb/\ub)$ are determined by the requirement that the 
expansion of $E({R})_p(\alpha,b;\ub,\vb)$ in powers of $\ub$ and $\vb$ 
gives $R(\ub,\vb)$ to order $p$. 
For each value of $\alpha$, $b$, and $p$, $E({R})_p(\alpha,b;\ub,\vb)$
provides an estimate of $R(\ub,\vb)$.

\begin{figure}[tb]
\centerline{\psfig{width=15truecm,angle=-90,file=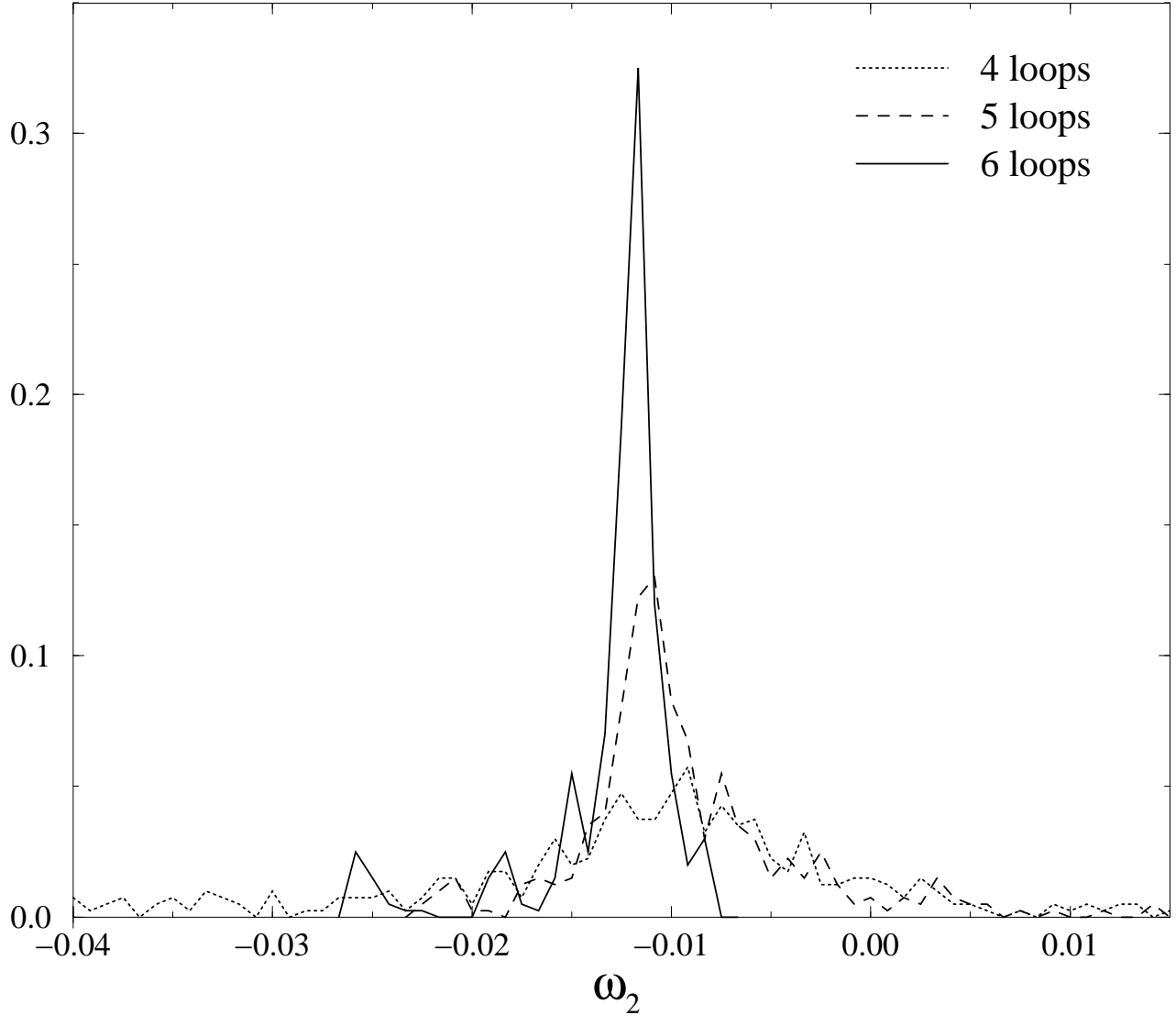}}
\caption{Distribution of the estimates of $\omega_2$ at the O(3)-symmetric 
fixed point.
}
\label{om2_symm}
\end{figure}

First of all, we have analyzed the stability properties of the $O(N)$-symmetric
fixed point. Since $\partial\beta_\vb/\partial \ub(\ub,0) = 0$, the 
eigenvalues are simply
\begin{equation}
\omega_1 =\, {\partial\beta_\ub\over \partial \ub}(\ub^*,0), \qquad\qquad
\omega_2 =\, {\partial\beta_\vb\over \partial \vb}(\ub^*,0),
\end{equation}
where $\ub^*$ is the fixed-point value of $\ub$. The exponent 
$\omega_1$ is the usual exponent that is considered in the 
$O(N)$-symmetric theory, while $\omega_2$ is the eigenvalue
that determines the stability of the fixed point.

In order to compute $\omega_2$, for many choices of the four parameters
$\alpha_1$, $b_1$, $\alpha_2$, and $b_2$, we have determined an estimate 
of $\ub^*$ and $\omega_2$ from the equations
\begin{eqnarray}
&& E(\beta_\ub/\ub)_p(\alpha_1,b_1;\ub^*_p(\alpha_1,b_1),0) = 0, 
\label{eq3.5} \\
&& \widehat{\omega}_2(\alpha_1,b_1,\alpha_2,b_2;p) = \, 
E(\partial\beta_\vb/\partial \vb)_p (\alpha_2,b_2;\ub^*_p(\alpha_1,b_1),0).
\end{eqnarray}
Note that $\ub^*_p(\alpha_1,b_1)$ is determined implicitly by
Eq. (\ref{eq3.5}), which has been solved numerically for each value of $\alpha_1$
and $b_1$. 

Then, we have considered sets of approximants such that 
$\alpha_{1,2}\in [\overline{\alpha} - \Delta\alpha,\overline{\alpha} +
\Delta\alpha]$ and 
$b_{1,2} \in [\overline{b} - \Delta b,\overline{b} + \Delta b]$. The final 
estimate was obtained averaging over all integer values of 
$\alpha_1$, $\alpha_2$, $b_1$, and $b_2$ belonging to these intervals.
The results for $N=3$ and several choices of the parameters are 
reported in Table \ref{critical_exponents_symm_N3_gexp}. 
In order to obtain a final estimate, we should devise a procedure 
to determine ``optimal" values for the parameters $\overline{\alpha}$,
$\overline{b}$, $\Delta\alpha$, and $\Delta b$. Reasonable values 
of $\overline{\alpha}$ and $\overline{b}$ can be obtained by
requiring that the estimates are approximately independent of the 
number of terms one is considering. It is less clear how to determine 
the width of the intervals $\Delta\alpha$ and $\Delta b$. 
Indeed, the results are stable, within the quoted errors, for many 
different choices of these two parameters, while the standard 
deviation of the estimate, which we use as an indication of the error, 
strongly depends on the choice one is making. In order to have 
reasonable error estimates, we have compared our results for 
$\ub^*$ with the estimates obtained from the analysis of the same series
by different authors. For $N=3$, Guida and 
Zinn-Justin quote $1.390(4)$ \cite{G-Z-98}.
We have therefore chosen our parameters 
so that we reproduce their errors. More precisely, we choose 
$\Delta\alpha=2$ and $\Delta b=3$, and quote the error in the final results as 
two standard deviations. With this choice, we obtain $\ub^* = 1.393(4)$,
which agrees with the previous estimate and has 
the same error. Results for other values of $N$ are reported in 
Table \ref{table_omega2_symmetricpoint}. Again, one can verify the good 
agreement of our estimates of $\ub^*$ with the results of Ref. \cite{G-Z-98}. 

In Table \ref{table_omega2_symmetricpoint} we also report our 
results for $\omega_2$. 
The $O(2)$-symmetric point is stable, since $\omega_2 = 0.103(8) > 0$.
On the other hand, the symmetric point is clearly unstable for $N\ge 4$.
For $N=3$, the analysis gives $\omega_2=-0.013(6) <0$, so that the fixed point
is unstable and $N_c < 3$. To better understand the reliability of the 
results, in Fig. \ref{om2_symm} we show the distribution of the 
estimates of $\omega_2$ when we vary $\alpha_1$, $b_1$,
$\alpha_2$, and $b_2$ in the chosen interval. 
It is evident that the quoted error on $\omega_2$ is quite conservative and 
that the results become increasingly stable as the number of loops increases. 

\begin{figure}[tb]
\centerline{\psfig{width=15truecm,angle=-90,file=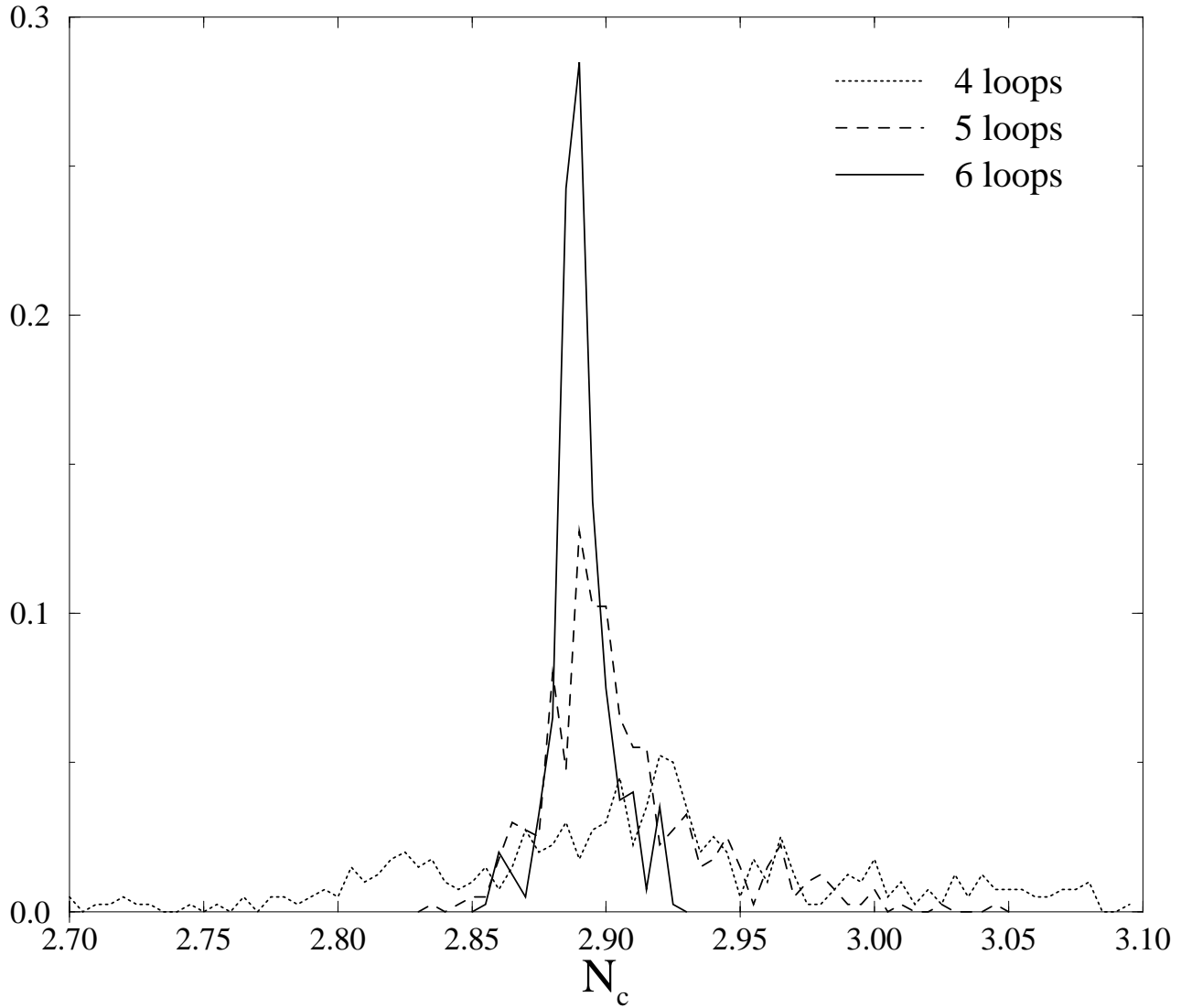}}
\caption{Distribution of the estimates of $N_c$. It is determined 
by requiring $\omega_2 = 0$ at the O($N$)-symmetric fixed point.
}
\label{nc_symm}
\end{figure}

We have then determined $N_c$, defined as the value of $N$ for which 
$\omega_2 = 0$ at the $O(N)$-symmetric fixed point 
\cite{footnote6}. The computation was done as before. For each value of the 
four parameters $\alpha_1$, $b_1$, $\alpha_2$, and $b_2$, we computed 
an estimate of $N_c$, by requiring
\begin{equation}
\widehat{\omega}_2(\alpha_1,b_1,\alpha_2,b_2;p) =\, 0.
\end{equation}
The distribution of the results is shown in Fig. \ref{nc_symm}.
It is evident that $N_c< 3$, as of course it should be expected from 
the analysis of $\omega_2$ at $N=3$. The result is increasingly stable
as the number of orders that are included increases. We estimate 
\begin{equation}
N_c = 2.91(9) \, \hbox{(\rm 4 loops)}, \qquad 
      2.91(3)\,  \hbox{(\rm 5 loops)}, \qquad
      2.89(2) \, \hbox{(\rm 6 loops)}.
\end{equation}
We can thus safely conclude that $N_c = 2.89(4)$.

\begin{table}[tbp]
%\squeezetable
\caption{
Estimates of the cubic fixed point and of the eigenvalues of the 
stability matrix for various values of $N$ and orders $p$ of the perturbative
series. Our final results correspond to $p= \hbox{\rm f}$.
}
\label{eigenvalues_cubicpoint}
\begin{tabular}{llllll}
$N$  & $p$  & $\ub^*$  &   $\vb^*$  & $\omega_1$  & $\omega_2$   \\
\hline
%% 3 & 4 & 1.328(16) & 0.094(20) & 0.753(11) & 0.022(6) \\
%%  & 5 & 1.328(12) & 0.087(16) & 0.792(10) & 0.003(4) \\
%%  & 6 & 1.315(4)  & 0.103(5)  & 0.776(5)  & 0.015(4) \\
%%  & f & 1.315(17) & 0.103(21) & 0.776(21) & 0.015(16)\\
%% \hline
% alpha in [0,4] b in [11-17] 
3 & 4 & 1.36(11)  & 0.04(13)  & 0.780(11) & 0.011(31)\\
  & 5 & 1.328(26) & 0.089(28) & 0.777(5)  & 0.009(6) \\
  & 6 & 1.321(9)  & 0.096(10) & 0.781(2)  & 0.010(2) \\
  & f & 1.321(18) & 0.096(20) & 0.781(4)  & 0.010(4) \\
\hline
%% 4 & 4 & 0.905(11) & 0.620(11) & 0.733(16) & 0.109(13) \\
%%   & 5 & 0.889(6)  & 0.632(5)  & 0.802(21) & 0.055(16) \\
%%   & 6 & 0.881(3)  & 0.639(2)  & 0.776(11) & 0.080(10) \\
%%   & f & 0.881(11) & 0.639(9)  & 0.776(37) & 0.080(35) \\
%% \hline
% alpha in [1/2,9/2] b on [8,14]
4 & 4 & 0.907(72) & 0.606(82)  & 0.711(77) & 0.144(78) \\
  & 5 & 0.883(17) & 0.639(17)  & 0.804(48) & 0.049(40) \\
  & 6 & 0.881(7)  & 0.639(7)   & 0.781(22) & 0.076(20) \\
  & f & 0.881(14) & 0.639(14)  & 0.781(44) & 0.076(40) \\
\hline
%% 8 & 4 & 0.460(10) & 1.133(4) & 0.747(14) & 0.157(7) \\
%%   & 5 & 0.447(5)  & 1.134(3) & 0.798(18) & 0.126(10)\\
%%   & 6 & 0.442(2)  & 1.135(2) & 0.782(8)  & 0.141(6) \\
%%  & f & 0.442(7)  & 1.135(3) & 0.782(24) & 0.141(21)\\
%% \hline
% alpha in [1,5] b in [7-13]
8 & 4 & 0.448(49) & 1.138(65) & 0.695(85) & 0.211(93) \\
  & 5 & 0.440(14) & 1.140(16) & 0.831(77) & 0.098(47) \\
  & 6 & 0.440(6)  & 1.136(5)  & 0.775(44) & 0.149(33) \\
  & f & 0.440(12) & 1.136(10) & 0.775(88) & 0.149(66) \\
\hline
% alpha in 
$\infty$ & 4 & 0.182(15) & 1.422(24)& 0.744(95) & 0.185(15) \\
         & 5 & 0.173(6)  & 1.424(7) & 0.783(31) & 0.177(6) \\
         & 6 & 0.174(3)  & 1.417(3) & 0.790(9)  & 0.178(3) \\
         & f & 0.174(6)  & 1.417(6) & 0.790(18) & 0.178(6) \\
%$\infty$ & 4 & 0.183(6)  & 1.423(10) & 0.778(18) & 0.186(5) \\
%         & 5 & 0.175(3)  & 1.420(3)  & 0.791(18) & 0.179(2) \\
%         & 6 & 0.174(2)  & 1.416(2)  & 0.787(6)  & 0.178(1) \\
%         & f & 0.174(3)  & 1.416(6)  & 0.787(10) & 0.178(2) \\
\end{tabular}
\end{table}

\begin{figure}[tb]
\centerline{\psfig{width=15truecm,angle=-90,file=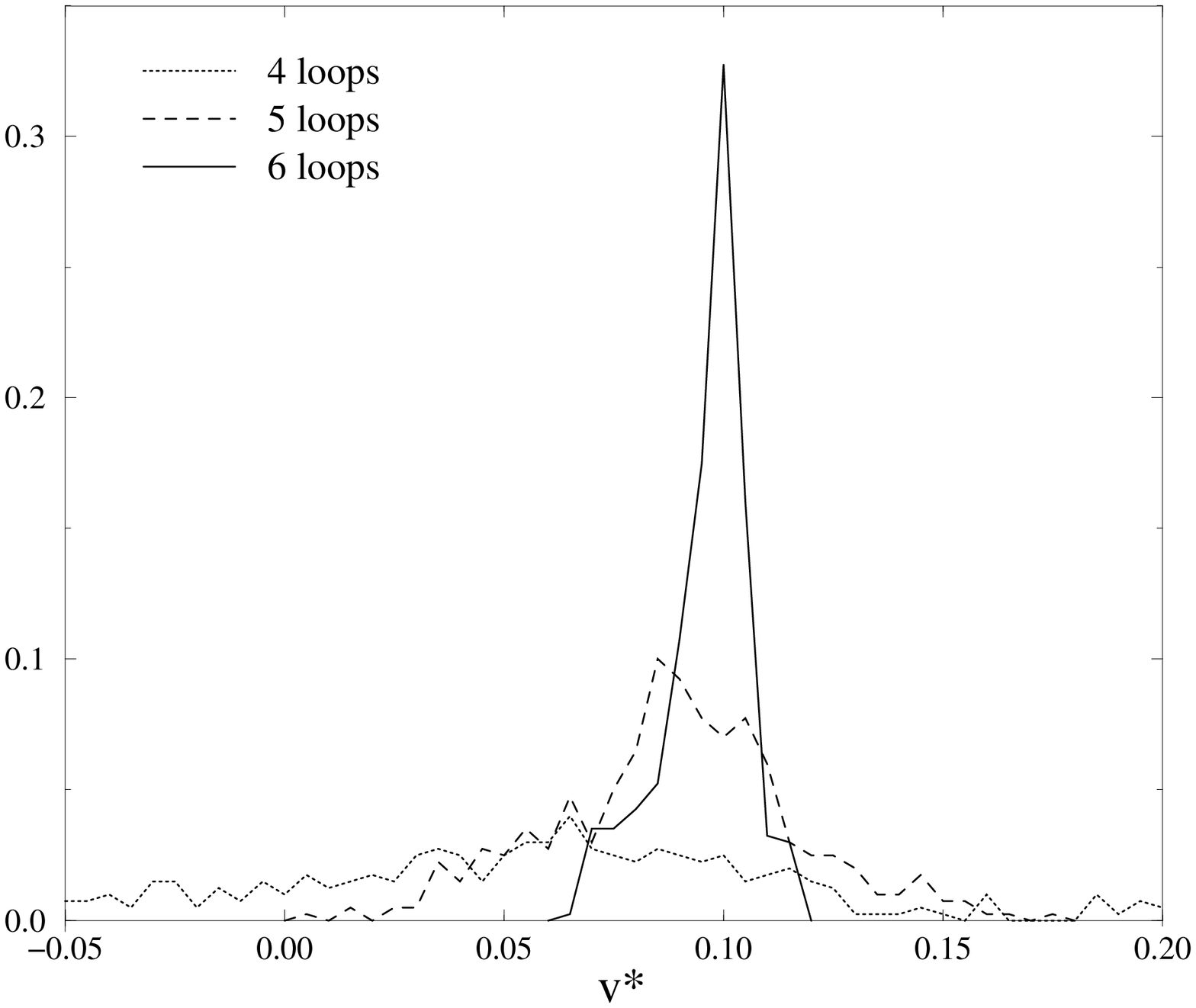}}
\caption{Distribution of 
$\bar{v}^*$ of the cubic fixed point for $N=3$.
}
\label{vc_h}
\end{figure}

\begin{figure}[tb]
\centerline{\psfig{width=15truecm,angle=-90,file=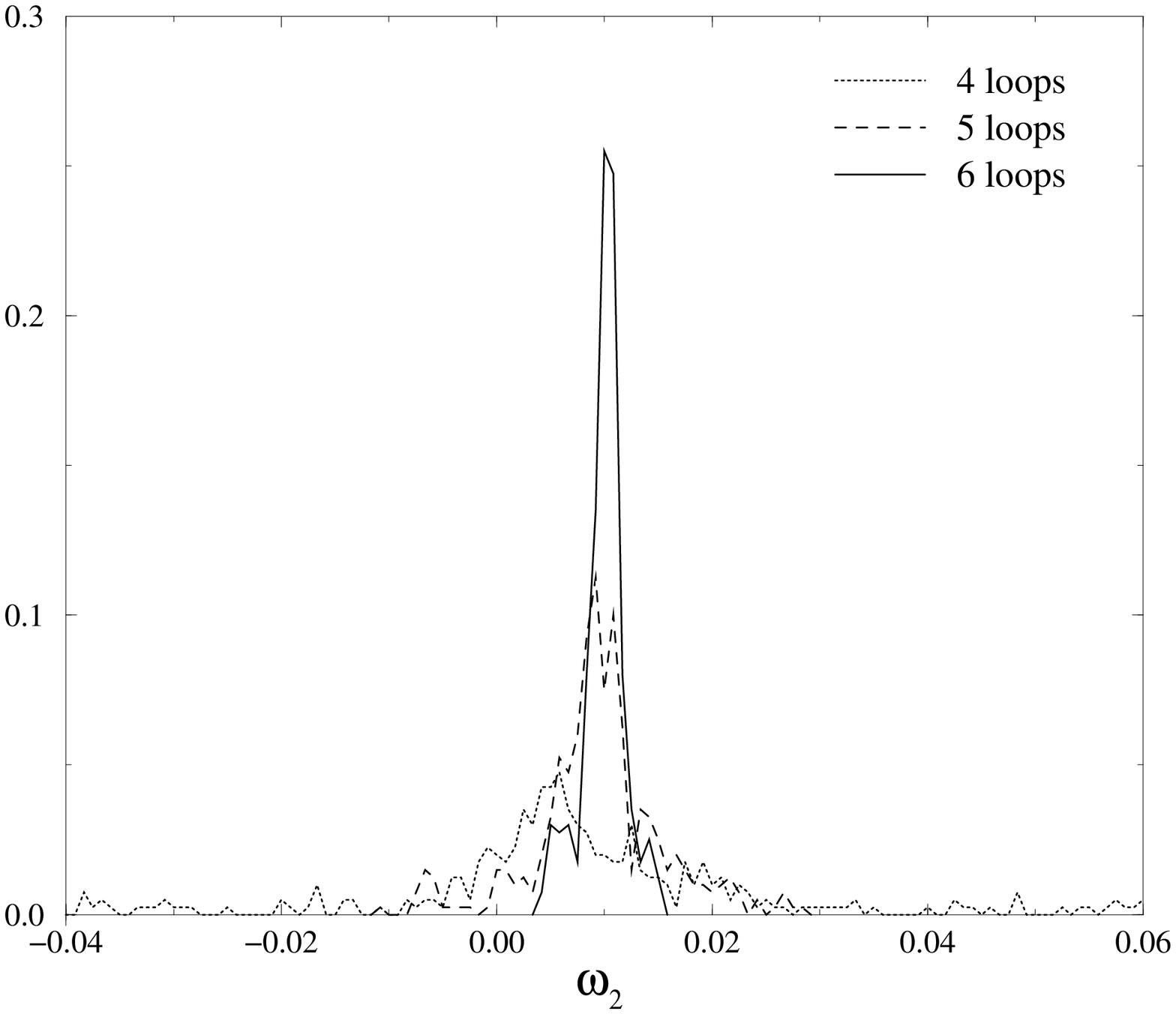}}
\caption{ Distribution of the estimates of $\omega_2$ at the cubic 
fixed point for $N=3$.
}
\label{om2_cub}
\end{figure}

We have then considered the cubic fixed point and studied its stability.
Again we used four parameters $\alpha_1$, $b_1$, $\alpha_2$, and $b_2$.
For each choice we first computed an estimate of the critical point
$\ub^*_p(\alpha_1,b_1,\alpha_2,b_2)$, $\vb^*_p(\alpha_1,b_1,\alpha_2,b_2)$ 
solving the equations
\begin{eqnarray}
E(\beta_\ub/\ub)_p(\alpha_1,b_1;\ub^*_p(\alpha_1,b_1,\alpha_2,b_2),
                         \vb^*_p(\alpha_1,b_1,\alpha_2,b_2)) &=& 0, \\
E(\beta_\vb/\vb)_p(\alpha_2,b_2;\ub^*_p(\alpha_1,b_1,\alpha_2,b_2),
                         \vb^*_p(\alpha_1,b_1,\alpha_2,b_2)) &=& 0.
\end{eqnarray}
Then we determined the elements of the stability matrix $\Omega$ from
\begin{eqnarray}
\Omega_{11} &=& E(\partial\beta_\ub/\partial \ub)_p
               (\alpha_1,b_1;\ub^*_p(\alpha_1,b_1,\alpha_2,b_2),
                            \vb^*_p(\alpha_1,b_1,\alpha_2,b_2)), \\
\Omega_{22} &=& E(\partial\beta_\vb/\partial \vb)_p
               (\alpha_2,b_2;\ub^*_p(\alpha_1,b_1,\alpha_2,b_2),
                            \vb^*_p(\alpha_1,b_1,\alpha_2,b_2)), \\
\Omega_{12} &=& \ub^*_p(\alpha_1,b_1,\alpha_2,b_2) \,
           E(1/\ub\,\cdot \partial\beta_\ub/\partial \vb)_{p-1}
               (\alpha_1,b_1;\ub^*_p(\alpha_1,b_1,\alpha_2,b_2),
                            \vb^*_p(\alpha_1,b_1,\alpha_2,b_2)), \\
\Omega_{21} &=& \vb^*_p(\alpha_1,b_1,\alpha_2,b_2) \,
           E(1/\vb\,\cdot \partial\beta_\vb/\partial \ub)_{p-1}
               (\alpha_2,b_2;\ub^*_p(\alpha_1,b_1,\alpha_2,b_2),
                            \vb^*_p(\alpha_1,b_1,\alpha_2,b_2)).
\end{eqnarray}
We computed the eigenvalues of $\Omega$, obtaining 
estimates of the exponents $\omega_1$ and $\omega_2$. 
As before, we determined $\overline\alpha$ and 
$\overline b$  by requiring the stability of the estimates of 
$\ub^*$, $\vb^*$, $\omega_1$, and $\omega_2$, when varying the order of the 
series. Errors were computed as before. The results are reported
in Table \ref{eigenvalues_cubicpoint}. They show that for 
$N\ge 3$ the cubic fixed point is stable. Note that
for $N=3$ we find $\vb^*>0$, which agrees 
with what is expected in the case of a stable cubic fixed point,
see Fig. \ref{rgflow}. 
The distribution of the estimates of $\bar{v}^*$, obtained varying
the parameters $\alpha_1, b_1, \alpha_2, b_2$, and reported in Fig.~\ref{vc_h},
shows that the result is quite stable. 
In accordance with this scenario, we find $\omega_2= 0.010(4) > 0 $. 
The quoted error is quite conservative, as it can be seen from 
Fig. \ref{om2_cub} in which we report the distribution of the estimates of $\omega_2$.
Note that the results become increasingly stable as the number of loops 
increases. 
One may compare our estimate of $\omega_2$ at the cubic fixed point with the 
four-loop result of Ref. \cite{Varnashev-99},  
$\omega_2 = 0.0081$, which is fully consistent with ours~\cite{footnoteVarnashev}.

\begin{table}
%\squeezetable
\caption{
Estimates of the critical exponents at the cubic fixed point
for various values of $N$ and order $p$ of the perturbative
series. The column ``final" reports our final results. The 
results of Ref. \protect\cite{Varnashev-99}, obtained by a 
Pad\'e-Borel analysis of the four-loop series,  are reported in the
last column.}
\label{critical_exponents_cubicpoint}
\begin{tabular}{lllllll}
$N$  &  &  $p = 4$  & $p = 5$ & $p = 6$ & final & 
                  Ref.~\protect\cite{Varnashev-99}\\
\hline
%% 3   & $\gamma$ & 1.385(12) & 1.392(9) & 1.386(4) & 1.386(10) \\
%%     & $\nu$    & 0.702(7)  & 0.707(5) & 0.704(2) & 0.704(5)  \\
%%     & $\eta$   & 0.0317(32)& 0.0328(17)& 0.0328(9) & 0.0328(9) \\
%%\hline 
 3   & $\gamma$ & 1.399(46) & 1.390(11)& 1.390(6) & 1.390(12) & 1.3775 \\
     & $\nu$    & 0.710(26) & 0.706(7) & 0.706(3) & 0.706(6)  & 0.6996 \\
     & $\eta$   & 0.0338(90)& 0.0325(20)& 0.0333(13) & 0.0333(26)& 0.0332 \\
%%\hline
%% 4   & $\gamma$ & 1.411(11) & 1.407(6) & 1.405(4) & 1.405(6)  \\
%%     & $\nu$    & 0.716(6)  & 0.714(4) & 0.714(2) & 0.714(2)  \\
%%     & $\eta$   & 0.0316(30)& 0.0311(12)& 0.0316(8) & 0.0316(13) \\
\hline
 4   & $\gamma$ & 1.415(41) & 1.403(12) & 1.405(5) & 1.405(10) & 1.4028 \\
     & $\nu$    & 0.718(23) & 0.714(7)  & 0.714(4) & 0.714(8)  & 0.7131 \\
     & $\eta$   & 0.0323(73)& 0.0305(18)& 0.0316(11) & 0.0316(22)& 0.0327 \\
\hline
%% 8   & $\gamma$ & 1.408(12) & 1.406(6) & 1.405(3) & 1.405(4)  \\
%%     & $\nu$    & 0.714(6)  & 0.714(4) & 0.713(2) & 0.713(3)  \\
%%     & $\eta$   & 0.0296(27)& 0.0298(10)& 0.0308(8) & 0.0308(18) \\
%% \hline
 8   & $\gamma$ & 1.401(38) & 1.402(12)& 1.404(5) & 1.404(10) & 1.4074 \\
     & $\nu$    & 0.709(21) & 0.711(7) & 0.712(3) & 0.712(6)  & 0.7153 \\
     & $\eta$   & 0.0292(61)& 0.0296(18)& 0.0306(10) & 0.0306(20)& 0.0324 \\
\hline
%% $\infty$
%%     & $\gamma$ & 1.401(12) & 1.396(6) & 1.396(3) & 1.396(3)  \\
%%     & $\nu$    & 0.710(7)  & 0.709(4) & 0.708(2) & 0.708(3)  \\
%%     & $\eta$   & 0.0287(29)& 0.0292(10)& 0.0304(8) & 0.0304(20) \\
$\infty$
     & $\gamma$ & 1.400(21) & 1.395(9) & 1.396(7) & 1.396(14)  & \\
     & $\nu$    & 0.709(12)  & 0.707(5) & 0.708(4) & 0.708(8)  & \\
     & $\eta$   & 0.0287(35)& 0.0294(11)& 0.0304(10) & 0.0304(20)& \\
\end{tabular}
\end{table}

We computed the exponents at the cubic fixed point, using 
Eqs. (\ref{eta_fromtheseries}), (\ref{nu_fromtheseries}), and 
(\ref{gamma_fromtheseries}). The results are reported in 
Table \ref{critical_exponents_cubicpoint}. 
Note that, for $N=3$, the exponents at the cubic fixed point
do not differ appreciably from those of the isotropic model. 
A recent reanalysis~\cite{G-Z-98} of the fixed-dimension expansion of the 
three-dimensional O($N$)-symmetric models obtained the following estimates for $N=3$:
$\gamma = 1.3895(50)$, $\nu=0.7073(35)$, and $\eta = 0.0355(25)$.
These results should be compared with the critical exponents at the cubic fixed-point:
$\gamma = 1.390(12)$, $\nu=0.706(6)$, and $\eta = 0.0333(26)$.
The difference is smaller than the precision of our results, 
which is about 1\% in the case of $\gamma$ and $\nu$.

We also checked the exact predictions for the exponents in the large-$N$
limit. For $N\to\infty$ with $\vb$ and $\ub$ fixed, 
we have $\beta_\vb(\ub,\vb) = \left. \beta (\vb)\right|_{\rm Ising}$,
so that $\overline{v}^* = \overline{v}^*_I$. 
Indeed, our estimate of $\vb^*$ is in 
agreement with the estimate obtained with the same method of analysis 
in Ref. \cite{B-L-Z-77}: $\vb^*_I = 1.416(5)$.
We have then compared our estimates $\eta=0.0304(20)$, 
$\nu=0.708(8)$, and $\gamma=1.396(14)$ with the 
predictions
\begin{eqnarray}
\eta &=& \eta_I = 0.0364(4), \\
\nu &=& {\nu_I\over (1 - \alpha_I)} = 0.7078(3), \nonumber \\
\gamma &=& {\gamma_I\over (1 - \alpha_I)} = 1.3899(7).\nonumber
\label{largeN-predictions}
\end{eqnarray}
The Ising model results have been taken from Ref. \cite{C-P-R-V-99-1}.
There is a substantial agreement,  although we note 
a small discrepancy for $\eta$. This small
difference is not unexpected. Indeed, 
the estimate of $\eta$ for the Ising model obtained from the fixed-dimension
expansion shows a systematic discrepancy with respect 
to high-temperature and Monte Carlo results. 
%
%% ANALISI VINCOLATA: non funziona !!!!!!
%
%% It was noted in 
%% Ref. \cite{C-P-R-V-99-1} that these discrepancies are reduced if one 
%% computes the critical exponents, fixing $\vb^*=1.402(2)$ which is 
%% the value predicted by the high-temperature analysis of 
%% Ref. \cite{C-P-R-V-99-1}. 
%% We have thus performed a constrained analysis, fixing $\vb^*=1.402(2)$
%% and computing $\ub^*$ from $\beta_\ub(\ub^*,\vb^*) = 0$. 
%% We obtained $\ub^* = 0.181(6)$, $\gamma = 1.407(14)$, $\nu = 0.714(6)$,
%% $\eta = 0.0314(32)$. The estimate of $\eta$ is in somewhat better 
%% agreement with the theoretical prediction, but the estimates of 
%% $\gamma$ and $\nu$ are clearly worse. We do not have any sound explanation.
%% Clearly, the discrepancy cannot 
%% be simply explained as the effect of the inaccurate estimate of $\vb^*$.

Finally, we checked the prediction $\omega_2 = -\alpha_I/\nu_I$ at the 
Ising fixed point. It is immediate to verify that 
\begin{equation}
\omega_2 = {\partial \beta_\ub\over \partial \ub} (0,\vb)
\end{equation}
is independent of $N$. 
Numerically we find 
\begin{equation}
\omega_2 = -0.174(31)\;         \hbox{(\rm 4 loops)}, \qquad 
\omega_2 = -0.178(7) \;         \hbox{(\rm 5 loops)},
\qquad \omega_2 = -0.177(3) \;  \hbox{(\rm 6 loops)}.
\end{equation}
Our final estimate is therefore $\omega_2 = -0.177(6)$ which should 
be compared with the prediction $\omega_2 = -\alpha_I/ \nu_I = -0.1745(12)$.

\section{Analysis of the $\epsilon$-expansion five-loop series}
\label{sec4}

In this Section we consider the alternative field-theoretic approach based 
on the $\epsilon$-expansion. 
The $\beta$-function and the exponents in 
the cubic model are known to $O(\epsilon^5)$~\cite{K-S-95}.
These series have already been the object of several analyses 
using different resummation methods
\cite{K-S-95,K-T-95,K-T-S-97,S-A-S-97,Varnashev-99}.
Ref.~\cite{K-T-S-97} studies the stability of the
$O(N)$-symmetric and of the cubic fixed points for  $N=3$. They rewrite 
the double expansion in $u$ and $v$ in terms of $g\equiv u+v$ and of the
parameter $\delta\equiv v/(u+v)$. Then, each coefficient in the expansion
in terms of $\delta$, which is a series in $g$, is resummed using the 
known large-order behavior. Since $v^*$, and hence $\delta^*$, is small,
one expects the resulting series to be rapidly convergent near the fixed 
point, and therefore
no additional resummation is applied~\cite{K-T-95}.
For the cubic fixed point,
Ref.~\cite{K-T-S-97} quotes 
$\omega_2\simeq 0.0049,\;0.0085,\;0.0021$, obtained respectively from the 
analysis of the three-, four-,
and five-loop series. Similar results (with the opposite sign) are found
for the O($N$)-symmetric fixed point. These estimates are compatible with 
a stable cubic fixed point, but the observed trend towards smaller 
values of $|\omega_2|$ leaves open the possibility that the estimate of 
$\omega_2$ 
will eventually change sign, modifying the conclusions about the stability 
of the fixed points. The same expansion has been analyzed in 
Ref.~\cite{S-A-S-97} using a Pad\'e-Borel resummation technique. The authors
report the estimate  $N_c\approx 2.86$, but again the
uncertainty on this result is not clear. 
We will present here a new analysis of the 
five-loop series using the method presented in the previous Section
and the value of the singularity of the Borel transform given below.

As before, we will consider expansions in $u$ with $z = v/u$ fixed.
The singularity closest to the origin of the Borel transform
is given by
\begin{eqnarray}
{1\over u_b} &= - \left( 1 + z \right) \qquad\qquad &{\rm for} \qquad\qquad z >0,
\label{bsingeps}\\
{1\over u_b} &= - \left( 1 + {1 \over  N}z \right)\qquad\qquad &{\rm for} \qquad\qquad z <0.\nonumber 
\end{eqnarray}
The fixed-point values
of $u$ and $z$ associated to the  O($N$)-symmetric and to the cubic fixed points
are respectively
\begin{equation}
u^* = {3\over 8+N} \epsilon + O(\epsilon^2),\qquad\qquad z^* = 0,
\label{onsim}
\end{equation}
and
\begin{equation}
u^* = {1\over N} \epsilon + O(\epsilon^2),\qquad\qquad 
z^* = {N-4\over 3} + O(\epsilon).
\label{cub}
\end{equation}
Since only the leading term in the $\epsilon$-expansion is relevant 
for the determination of the singularity 
$\epsilon_b$ \cite{Lipatov-77,B-L-Z-77},
using Eqs. (\ref{onsim}), (\ref{cub}), and (\ref{bsingeps}), we find 
\begin{equation}
\epsilon_b = -{N+8\over 3}
\label{epsbs}
\end{equation}
for the O($N$)-invariant fixed point~\cite{Lipatov-77,B-L-Z-77},
and
\begin{eqnarray}
\epsilon_b = & -{3 N^2 \over 4(N-1)}\qquad\qquad &
                  {\rm for}\qquad\qquad N<4,
\nonumber \\
\epsilon_b = & -{3 N \over N-1}\qquad\qquad      &
                  {\rm for}\qquad\qquad N>4, 
\label{epsbc}
\end{eqnarray}
for the cubic fixed point.

\begin{table}
%\squeezetable
\caption{
$\epsilon$-expansion estimates of the critical exponents 
$\omega_1$ and $\omega_2$ 
at the cubic and at the $O(N)$-symmetric fixed points for $N=3$.
The number $p$ indicates the number of loops included in the analysis,
the columns labelled by $\alpha$ and $b$ indicate the intervals 
of $\alpha$ and $b$ used, ``final" indicates our
final estimate from the given analysis.}
\label{critical_exponents_N3_eps}
\begin{tabular}{lccllll}
& $\alpha$ &  $b$   &   $p=3$   &    $p=4$  &   $p=5$   & final \\
\hline
\multicolumn{7}{c}{$O(N)$-symmetric fixed point} \\
\hline
$\omega_1$ & [$-$0.5,1.5] & [11,17] &
         0.799(20)  &  0.790(8) & 0.795(4) & 0.795(8) \\
           & [$-$0.5,1.5] & [8,20]  &
         0.804(36)  &  0.784(20)& 0.801(14)& 0.801(28)\\
%           & [$-$0.5,2.0] & [8,20]  &
%         0.820(55)  &  0.782(29)& 0.802(19)& 0.802(39)\\
           & [$-$1.5,2.5] & [11,17] &
         0.830(98)  &  0.784(19)& 0.795(10)& 0.795(20)\\
\hline
$\omega_2$ & [0.0,2.0]  & [12,18]  &
         0.004(6)   &  0.000(3)   & $-$0.003(2) & $-$0.003(4) \\
           & [0.0,2.0]  & [9,21]  &
         0.005(9)   & $-$0.002(5) & $-$0.003(2) & $-$0.003(4) \\
%           & [$-$0.5,2.0] & [8,20]  &
%         0.008(10)  & $-$0.002(7) & $-$0.002(3) & $-$0.002(3) \\
           & [$-$1.0,3.0] & [12,18]  &
         0.022(32)  &    0.000(7) & $-$0.003(4) & $-$0.003(8) \\ 
\hline\hline
\multicolumn{7}{c}{cubic fixed point} \\
\hline
$\omega_1$ 
%%         & [$-$0.5,1.5] & [10,16] &
%%         0.803(23)  &  0.792(9) & 0.799(5) & 0.799(14) \\
%%           & [$-$0.5,1.5] & [7,19] &
%%         0.810(41)  &  0.784(24)& 0.805(18)& 0.805(39) \\
%           & [$-$0.5,2.0] & [7,19] &
%         0.827(61)  &  0.782(35)& 0.808(25)& 0.808(51) \\
%%            & [$-$1.5,2.5] & [10,16] &
%%         0.84(10)    &  0.785(22)& 0.798(12)& 0.798(25) \\
          & $[-0.5,1.5]$ & [12,18] &
         0.789(16)   &  0.797(6) & 0.796(2) & 0.796(4)  \\
          & $[-0.5,1.5]$ & [9,21] &
         0.793(31)& 0.793(13) & 0.799(7)  & 0.799(14) \\
          & $[-1.5,2.5]$ & [12,18] &
         0.82(10)    &  0.794(20)& 0.793(6) & 0.793(12) \\
\hline
$\omega_2$ & [$-$0.5,1.5] & [11,17] &
         0.005(12)  &  0.006(4) & 0.007(2) & 0.007(4) \\
           & [$-$0.5,1.5] & [8,20]  &
         0.004(13)  &  0.006(4) & 0.006(2) & 0.006(4) \\
%           & [$-$0.5,2.0] & [8,20]  &
%         0.004(13)  &  0.006(5) & 0.006(2) & 0.006(2) \\
          & [$-$1.5,2.5] & [11,17] &
        $-$0.004(17)&  0.002(7) & 0.006(3) & 0.006(6) \\
\end{tabular}
\end{table}

\begin{table}
%\squeezetable
\caption{
Estimates of the critical exponent $\omega_2$ 
at the $O(N)$-symmetric fixed point for several values of $N$.
The number $p$ indicates the number of loops included in the analysis,
the columns labelled by $\alpha$ and $b$ indicate the interval 
of $\alpha$ and $b$ used, ``final" reports our final estimate.}
\label{omega_ON_eps}
\begin{tabular}{lccllll}
$N$ & $\alpha$ &  $b$   &   $p=3$   &    $p=4$  &   $p=5$   & final \\
\hline
%% 2   & $[-1.0,3.0]$ & [10,16] &
%%            0.125(21)  & 0.115(4) & 0.114(2) & 0.114(4) \\
2  & $[0.0,2.0]$ & [9,21] &
           0.116(12)  & 0.113(5) & 0.114(2) & 0.114(4) \\
%% 3   & $[-1.0,3.0]$ & [12,18] &
%%           0.022(32)  & 0.000(7) & $-$0.003(4) & $-$0.003(8) \\
3   & [0.0,2.0]  & [9,21]  &
         0.005(9)   & $-$0.002(5) & $-$0.003(2) & $-$0.003(4) \\
%% 4   & $[-0.5,3.5]$ & [12,18] &
%%         $-$0.070(35)  & $-$0.103(11) & $-$0.105(5) & $-$0.105(10) \\
4   & $[0.5,2.5]$ & [8,20] &
        $-$0.088(11)  & $-$0.103(7)  & $-$0.105(3) & $-$0.105(6) \\
%% 8   & $[0.5,4.5] $ & [10,16] &
%%        $-$0.393(60)   & $-$0.394(16)& $-$0.395(5)& $-$0.395(10) \\
8   & $[1.5,3.5] $ & [7,19] &
        $-$0.374(24)   & $-$0.395(12)& $-$0.395(4)& $-$0.395(8) \\
\end{tabular}
\end{table}

The method of analysis is the one  explained in the previous Section
with two simplifications: first, we consider expansions in one variable 
only; moreover, the value at which we should compute the expansion is known 
($\epsilon=1$). Each series is resummed using several different values of 
$\alpha$ and $b$ belonging to the intervals 
$\alpha\in[\overline{\alpha}-\Delta\alpha,\overline{\alpha}+\Delta\alpha]$
and $b\in[\overline{b}-\Delta b, \overline{b} +\Delta b]$. The parameters
$\overline{\alpha}$ and $\overline{b}$ were chosen as before, while, after 
several trials, we fixed $\Delta\alpha=1$ and $\Delta b = 6$. The error in 
the final results is always two standard deviations. The dependence of the 
results on the different choices of $\alpha$ and $b$ is shown in 
Table \ref{critical_exponents_N3_eps}. The final results are reported 
in Tables \ref{omega_ON_eps} and \ref{omega_cubic_eps}. 

The final results are in reasonable agreement with the estimates of the 
previous Section. However,
the instability of the $O(N)$-symmetric fixed point is less clear 
in the $\epsilon$-expansion: indeed we find $\omega_2 = -0.003(4)$ 
at the symmetric fixed point and $\omega_2 = 0.006(4)$ at the cubic fixed 
point. This is not surprising: for the $O(N)$-symmetric model 
the five-loop $\epsilon$-expansion gives results that are less precise 
than the six-loop expansion in fixed dimension. 
We also mention that for $N=\infty$ 
we obtain an estimate of $\eta$, $\eta=0.0349(22)$, that is in 
perfect agreement with the theoretical prediction $\eta=0.0364(4)$. 

\begin{table}
%\squeezetable
\caption{
Estimates of the critical exponents 
at the cubic fixed point for several values of $N$.
The number $p$ indicates the number of loops included in the analysis,
the columns labelled by $\alpha$ and $b$ indicate the intervals 
of $\alpha$ and $b$ used, ``final" reports our
final estimate.}
\label{omega_cubic_eps}
\begin{tabular}{lccllll}
$N$ & $\alpha$ &  $b$   &   $p=3$   &    $p=4$  &   $p=5$   & final \\
\hline
\multicolumn{7}{c}{$\omega_1$} \\
\hline
%% 3 & $[-1.5,3.5]$ & [10,16] &
%%       0.84(10) & 0.785(22) & 0.798(12) & 0.798(25) \\
3 & $[-0.5,1.5]$ & [9,21] &
       0.793(31)& 0.793(13) & 0.799(7)  & 0.799(14) \\
%% 4 & $[-2.0,2.0]$ & [13,19] &
%%        0.78(5) & 0.784(16) & 0.785(6) & 0.785(7) \\
4 & $[-1.0,1.0]$ & [10,22] &
       0.790(20) & 0.787(8) & 0.790(4) & 0.790(8) \\
%% 8 & $[-2.0,2.0]$ & [12,18] &
%%        0.76(3) & 0.780(19) & 0.782(7) & 0.782(9) \\
8 & $[-1.0,1.0]$ & [8,20] &
       0.779(16) & 0.782(5) & 0.786(3) & 0.786(6) \\
%% $\infty$ & $[-2.0,2.0]$ & [12,18] &
%%       0.77(6) & 0.787(31) & 0.798(17)& 0.798(28) \\
$\infty$ & $[-1.0,1.0]$ & [10,22] &
       0.772(22) & 0.789(11) & 0.802(9)& 0.802(18) \\
\hline
\multicolumn{7}{c}{$\omega_2$} \\
\hline
%% 3 & $[-1.5,3.5]$ & [11,17] &
%%      $-$0.004(17) & 0.002(7) & 0.006(3) & 0.006(7) \\
3 & $[-0.5,1.5]$ & [8,20] &
        0.004(13) & 0.006(4) & 0.006(2) & 0.006(4) \\
%% 4 & $[-2.0,2.0]$ & [14,20] & 
%%         0.024(84) & 0.083(16)& 0.081(4) & 0.081(6) \\
4 & $[-1.5,0.5]$ & [11,23] & 
        0.073(15) & 0.078(5) & 0.078(2) & 0.078(4) \\
%% 8 & $[-2.5,1.5]$ & [13,19] &
%%         0.149(14) & 0.152(7) & 0.154(3) & 0.154(5) \\
8 & $[-1.5,0.5]$ & [11,23] &
        0.154(9) & 0.155(4) & 0.155(1) & 0.155(2) \\
%% $\infty$ & $[-2.5,1.5]$ & [13,19] &
%%         0.204(19) & 0.201(10)& 0.195(8) & 0.195(14) \\
$\infty$ & $[-2.5,-0.5]$ & [11,23] &
        0.210(17) & 0.208(6)& 0.202(4) & 0.202(8) \\
\hline
\multicolumn{7}{c}{$\gamma$} \\
\hline
%% 3 & $[0.5,4.5]$ & [3,9] &
%%      1.39(11) & 1.377(13) & 1.376(5) & 1.376(6) \\
3 & $[1.5,3.5]$ & [3,15] &
     1.370(29)& 1.375(8)  & 1.377(3) & 1.377(6) \\
%% 4 & $[0.5,4.5]$ & [4,10] &
%%      1.43(9)  & 1.420(11) & 1.417(4) & 1.417(7) \\
4 & $[1.5,3.5]$ & [3,15] &
     1.414(20)  & 1.421(7) & 1.419(3) & 1.419(6) \\
%% 8 & $[-1.5,3.5]$ & [5,11] &
%%      1.41(4)  & 1.418(13) & 1.419(5) & 1.419(6) \\
8 & $[0.5,2.5]$ & [3,15] &
     1.420(22)  & 1.424(8) & 1.422(3) & 1.422(6) \\
%% $\infty$ & $[-1.5,3.5]$ & [6,12] &
%%      1.40(3)  & 1.397(11) & 1.397(4) & 1.397(4) \\ 
$\infty$ & $[0.0,2.0]$ & [3,15] &
    1.394(27) & 1.399(11) & 1.399(4) & 1.399(8) \\
\hline
\multicolumn{7}{c}{$\nu$} \\
\hline
%% 3 & $[0.5,4.5]$ & [3,9] &
%%      0.707(60) & 0.701(8) & 0.701(3)  & 0.701(3) \\
3 & $[1.5,3.5]$ & [3,15] &
     0.695(16) & 0.699(5) & 0.701(2)  & 0.701(4) \\
%% 4 & $[0.5,4.5]$ & [4,10] &
%%      0.728(47)  & 0.723(7) & 0.722(2) & 0.722(3)\\
4 & $[1.5,3.5]$ & [3,15] &
     0.717(11)  & 0.723(5) & 0.723(2) & 0.723(4)\\
%% 8 & $[0.0,4.0]$ & [3,9] &
%%      0.725(20)  & 0.722(4) & 0.722(1) & 0.722(1) \\
8 & $[1.0,3.0]$ & [3,15] &
     0.723(7)  & 0.724(3) & 0.723(1) & 0.723(2) \\
%% $\infty$ & $[0.0,4.0]$ & [3,9] &
%%      0.718(25)  & 0.709(3) & 0.711(1) & 0.711(3) \\
$\infty$ & $[1.0,3.0]$ & [3,15] &
     0.713(4)  & 0.711(2) & 0.711(1) & 0.711(2) \\
\hline
\multicolumn{7}{c}{$\eta$} \\
\hline
%% 3 & $[0.0,4.0]$ & [4,10] &
%%      0.0311(73) & 0.0353(20) & 0.0378(16) & 0.0378(41) \\
3 & $[1.0,3.0]$ & [3,15] &
     0.0319(61) & 0.0359(16) & 0.0374(11) & 0.0374(22) \\
%% 4 & $[0.0,4.0]$ & [4,10] &
%%      0.0321(51) & 0.0343(13) & 0.0371(14) & 0.0371(42) \\
4 & $[0.5,2.5]$ & [4,16] &
     0.0319(23) & 0.0339(10) & 0.0357(9)  & 0.0357(18) \\
%% 8 & $[0.0,4.0]$ & [3,9] &
%%      0.0295(61) & 0.0340(19) & 0.0357(11) & 0.0357(28) \\
8 & $[0.5,2.5]$ & [3,15] &
     0.0301(31) & 0.0336(11) & 0.0349(8) & 0.0349(16) \\
%% $\infty$ & $[-0.5,3.5]$ & [3,9] &
%%      0.0306(42) & 0.0325(15) & 0.0356(18) & 0.0356(49) \\
$\infty$ & $[0.5,2.5]$ & [3,15] &
     0.0296(36) & 0.0332(14) & 0.0349(11) & 0.0349(22) \\
\end{tabular}
\end{table}

Finally we compute $N_c$. We will determine it following two different 
strategies. A first possibility is to consider $\omega_2$ and determine the 
value of $N$ for which the three-dimensional estimate of $\omega_2$ 
vanishes, which  is exactly what we did in the fixed-dimension 
expansion. At the symmetric fixed point  we find
\begin{equation}
N_c = 3.07(9) \, \hbox{(\rm 3 loops)}, \qquad 
      2.99(6)\,  \hbox{(\rm 4 loops)}, \qquad
      2.98(2) \, \hbox{(\rm 5 loops)},
\end{equation}
while at the cubic fixed point 
\begin{equation}
N_c = 3.04(10) \, \hbox{(\rm 3 loops)}, \qquad 
      2.95(6)\,  \hbox{(\rm 4 loops)}, \qquad
      2.96(3) \, \hbox{(\rm 5 loops)}.
\end{equation}
Averaging the two results and reporting the error as two standard deviations,
we obtain $N_c = 2.97(6)$, which is in agreement with the analysis of the 
same series of Ref. \cite{K-S-95}, $N_c \approx 2.958$. 
It is also consistent with the results of the analysis of the eigenvalues 
$\omega_2$ for 
$N=3$: indeed we found $\omega_2 > 0$ 
and $\omega_2 < 0$ at the cubic and at the symmetric point respectively, 
implying $N_c < 3$. However, 
the error bars on $\omega_2$ are such that the opposite inequalities are not 
excluded. Analogously here, although $N_c < 3$, the error
is such that it does not exclude $N=3$. In practice, from this  
analysis alone, it is impossible to conclude 
safely that $N_c < 3$.

A second possibility consists in solving the equation $\omega_2=0$ 
perturbatively, obtaining for $N_c$ an expansion in powers of 
$\epsilon$. The result is independent of which fixed point is chosen. 
Unfortunately, the singularity of the Borel transform of 
$N_c(\epsilon)$ is not known and therefore, we used the 
Pad\'e-Borel method. The analysis of the 
series already appears in Ref. \cite{Varnashev-99}, and will not be repeated 
here. Instead, we will try to make use of the fact that $N_c=2$ for 
$d=2$, performing a constrained analysis. The method has already 
been applied in many instances
\cite{L-Z-87,C-P-R-V-98,P-V-gr-98,Pelissetto-Vicari-effpot,%
Pelissetto-Vicari_00}, providing more precise estimates of the critical 
quantities. The method consists in rewriting 
\begin{equation}
N_c(\epsilon) = 2 + (2 - \epsilon) \Delta(\epsilon),
\end{equation}
where
\begin{equation}
\Delta(\epsilon) = {N_c(\epsilon) - 2\over (2 - \epsilon)}.
\end{equation}
The quantity $\Delta(\epsilon)$ is expanded in powers of $\epsilon$ and 
then resummed. One can verify that the coefficients of the expansion 
of $\Delta(\epsilon)$ are uniformly smaller than the coefficients of the 
original series, by a factor of two approximately. Therefore,
one expects a corresponding improvement in the error estimates. 
We considered Pad\'e's [2/1], [1/2], [2/2], and [3/1] and several different 
values of the parameter $b$ between the ``reasonable" values 0 and 20.
Pad\'e's [1/2] and [2/2] have a singularity on the real positive axis 
for $b\lesssim 7$: these cases are of course excluded from consideration.
At four loops, we find that the approximant [2/1] gives estimates 
varying between 2.82 and 2.87, while the approximant [1/2] is stable 
giving $N_c\approx 2.82$. At five loops, the approximant [3/1] varies between 
2.85 and 2.92, while the approximant [2/2] is more stable and gives 
$N_c\approx 2.83$. To appreciate the improvement of the results due to the 
constrained analysis, we report the corresponding variation of the 
estimates for the original series: for $0\le b \le 20$ the approximants 
[2/1], [1/2], [2/2], and [3/1] give estimates varying in the intervals 
$2.80\le N_c\le 2.92$, $2.90\le N_c \le 2.96$, 
$2.90\le N_c\le 2.95$, $2.86\le N_c \le 3.03$. 
A conservative final estimate is 
$N_c = 2.87(5)$. This result is lower than that of the previous analysis,
but still compatible with it. 
The constrained analysis is in much better agreement with the results obtained
in the fixed-dimension expansion, and clearly supports the claim that 
$N_c < 3$.

\acknowledgements
We thank Michele Caselle, Martin Hasenbusch, Victor Mart\'\i n-Mayor, 
Giorgio Parisi, and Paolo Rossi for useful discussions.
One of us (J.M.C.) acknowledges support from the EC TMR program
ERBFMRX-CT97-0122.

\appendix

\section{Cubic fixed point in the $\epsilon$-expansion}

We report here the $\epsilon$-expansion computation of the 
four-point renormalized couplings $\ub^*$ and
$\vb^*$ defined in Sec. \ref{sec2} at the cubic fixed point,
following Refs. \cite{P-V-gr-98,Pelissetto-Vicari_00}. 
We have calculated the three-loop expansion of these two quantities.
The final result can be written as 
\begin{equation}
\bar u^*(\epsilon)=\, {(N+8)\over 3N} \sum_{n=0}\bar u_n\epsilon^n \; ,
     \quad \quad \quad 
\bar v^*(\epsilon)=\,  \sum_{n=0}\bar v_n\epsilon^n\; ,
\end{equation}
where
\begin{eqnarray}
&&\bar u_0 =  1, \\ 
&&\bar u_1 = - {(N-1)(19 N - 106)\over 27 N^2} \; , \\
&& \bar{u}_2 = \,
   - {\frac{2\,\left( -11236 + 22540\,N -
         14181\,{N^2} + 2446\,{N^3} +
         107\,{N^4} \right) }{729\,{N^4}}} 
\nonumber \\
&& \quad 
  - {\frac{ \lambda\,\left( N -1 \right) \,
         \left( 86 + 7\,N \right)  }{81\,
       {N^2}}} 
   + {\frac{4\,\left( 14 - 7\,N - 6\,{N^2} +
         2\,{N^3} \right) \,\zeta(3)}{9\,
       {N^3}}} \; ,
\\ \nonumber \\
&& \bar{u}_3 =\, 
  {\frac{8\,H\,\left( 10 - 5\,N - 2\,{N^2}
          \right) }{27\,{N^3}}} +
   {\frac{\lambda\,\left( -9116 + 16328\,N -
         7863\,{N^2} + 209\,{N^3} +
         37\,{N^4} \right) }{729\,{N^4}}}
\nonumber \\
&&  \quad - {1\over 157464\,{N^6}} 
    \left(47640640 - 134959200\,N +
       141439956\,{N^2} - 64950380\,{N^3} \right. 
\nonumber \\
&& \qquad \quad \left. +
       11140557\,{N^4} - 136770\,{N^5} +
       11821\,{N^6}\right)
\nonumber \\
&& \quad + {\frac{\left( 16 - 8\,N - 8\,{N^2} +
         3\,{N^3} \right) \,{{\pi }^4}}{405\,
       {N^3}}} -
   {\frac{\left( N-1 \right) \,
       \left( 86 + 7\,N \right) \,
       \left( \gamma_E\,\lambda + Q_1 \right) }{162\,
       {N^2}}} 
\nonumber \\ 
&& \quad - {\frac{4\,\left( -68 + 54\,N - 4\,{N^2} +
         3\,{N^3} \right) \,Q_2}{27\,
       {N^3}}} 
   - {\frac{40\,\left( 18 - 5\,N - 2\,{N^2} -
         6\,{N^3} + 2\,{N^4} \right) \,
       \zeta(5)}{27\,{N^4}}}
\nonumber \\
&& \quad + {\frac{\left( -59360 + 93392\,N -
         27220\,{N^2} - 14140\,{N^3} +
         5083\,{N^4} + 346\,{N^5} \right) \,
       \zeta(3)}{486\,{N^5}}} \; ,
\end{eqnarray}
and
\begin{eqnarray}
&&\bar v_0 = \frac{N - 4}{N}\; , \\ 
\nonumber \\
&& \bar v_1 = {(N-1)(-424 + 110 N + 17 N^2)\over 27 N^3}\; , \\
\nonumber \\
&& \bar{v}_2 =\, 
  - {\frac{25\,\lambda\,\left( N-4 \right) \,
       \left( N-1 \right) \,\left( 2 + N \right)
       }{81\,{N^3}}} 
   - {\frac{4\,\left( 2 + N \right) \,
       \left( 28 - 32\,N + 6\,{N^2} +
         {N^3} \right) \,\zeta(3)}{9\,
       {N^4}}}
\nonumber \\
&& \quad + {\frac{2\,\left( -44944 + 93764\,N -
         61408\,{N^2} + 10951\,{N^3} +
         565\,{N^4} + 100\,{N^5} \right) }{729\,
       {N^5}}} \; ,
\\ \nonumber \\
&& \bar{v}_3 =\, 
  {\frac{8\,H\,\left( N-4 \right) \,
       \left( 8 - 4\,N - {N^3} \right) }{27\,
       {N^4}}} 
\nonumber \\ 
&& \quad -
   {\frac{\lambda\,\left( -21200 + 31540\,N -
         6536\,{N^2} - 5953\,{N^3} +
         581\,{N^4} + 353\,{N^5} \right) }{729\,
       {N^5}}} 
\nonumber \\
&& \quad + {1\over 157464\, N^7} \left(
   190562560 - 555117760\,N +
       598171440\,{N^2} - 281214884\,{N^3} \right. 
\nonumber \\ 
&& \qquad \qquad \left. +
       47944832\,{N^4} + 267603\,{N^5} -
       77234\,{N^6} + 23315\,{N^7} \right)
\nonumber \\
&& \quad    - {\frac{\left( 68 - 42\,N - 32\,{N^2} +
         14\,{N^3} + {N^4} \right) \,{{\pi }^4}}
       {405\,{N^4}}} -
   {\frac{25\,\left( N-4 \right) \,
       \left( N-1  \right) \,
       \left( 2 + N \right) \,
       \left( \gamma_E\,\lambda + Q_1 \right) }{162\,
       {N^3}}} 
\nonumber \\ 
&& \quad - {\frac{4\,\left( N-4 \right) \,
       \left( -24 + 22\,N - 5\,{N^2} +
         2\,{N^3} \right) \,Q_2}{9\,
       {N^4}}} 
\nonumber \\
&& \quad 
   + {\frac{40\,\left( 72 - 30\,N - 11\,{N^2} -
         18\,{N^3} + 7\,{N^4} +
         {N^5} \right) \,\zeta(5)}{27\,
       {N^5}}}
\nonumber \\
&&   \quad  - {\frac{\left( -237440 + 401536\,N -
         151152\,{N^2} - 36648\,{N^3} +
         15662\,{N^4} + 2277\,{N^5} +
         68\,{N^6} \right) \,\zeta(3)}{486\,
       {N^6}}}\; . 
\nonumber \\ 
{}
\end{eqnarray}
Here $\gamma_E$ is Euler's constant. We have also introduced the 
following numerical constants \cite{Pelissetto-Vicari_00}:
\begin{eqnarray}
\lambda &=&\hphantom{-} 1.171 953 619 344 729 445... \nonumber \\
Q_1     &=&          -  2.695 258 053 506 736 953... \nonumber \\
Q_2     &=&\hphantom{-} 0.400 685 634 386 531 428... \nonumber \\
H       &=&          -  2.155 952 487 340 794 361...
\end{eqnarray}
A standard analysis \cite{P-V-gr-98} gives:
\begin{eqnarray}
N = 3: && \qquad \ub^* =\, 1.416(10), \quad\qquad \vb^* =\, -0.03(14)\; ; 
\\
N = 4: && \qquad \ub^* =\, 0.971(19), \quad\qquad \vb^* =\, 0.58(9)\; ; 
\\
N = 8: && \qquad \ub^* =\, 0.455(73), \quad\qquad \vb^* =\, 1.14(9).
\end{eqnarray}
The estimates of $\vb^*$ are in reasonable agreement with the results
of Section \ref{sec3}, although they are much less precise, as it should 
be expected since here we are analyzing a shorter series. 
On the other hand, the estimates of $\ub^*$ strongly 
disagree with the quoted error bars. The estimate for $N = 3$ is the 
worst one: indeed, the six-loop fixed-dimension expansion predicts
$\ub^* =\, 1.321(18)$. However, the quoted errors (based, as usual,
on the stability of the estimates when changing the parameters
$b$ and $\alpha$) seem to be largely underestimated.
For instance, the three-loop series for $\ub^*$ at the $O(N)$-symmetric 
fixed point gives $\ub^* = 1.39(7)$ 
\cite{Pelissetto-Vicari_00}: the error is in this case seven times larger!
There is no reason to believe the error on the estimate of $\ub^*$ 
at the cubic point to be much smaller than that at the isotropic one. 
Moreover, it is difficult to accept that an expansion truncated at three 
loops might give a more precise result than the six-loop fixed-dimension 
expansion.
Thus, the previous error estimates should not be trusted and the observed
stability is probably accidental.
If we assume that the correct error is of order $\approx 0.07$ as in the 
$O(3)$ case, then all results are in agreement.
Note that the errors on $\vb^*$ are instead of the expected order of magnitude.

% ========================= REFERENCES =========================


\begin{references}
\bibitem{footnote1}
Renormalization-group trajectories leading to
unstable regions have been considered in the study of
fluctuation-induced first-order transitions
(see e.g. Refs.~\cite{fifopt}).
\bibitem{footnote2}
It should be noted that, for $v_0\to -\infty$, one obtains the model 
described in Refs. \cite{Kim-etal_75,Kim-Levy_75} in which the spins 
align along the lattice axes. A high-temperature analysis on the 
fcc lattice indicates that these models have a first-order transition
for $N>2.35\pm0.20$. This is consistent with our argument that 
predicts the transition to be of first order for any $v_0<0$ and 
$N>N_c$. More general models that have Eq. (\ref{Hphi4}) as their
continuous spin limit for $v_0\to -\infty$ have also
been considered in Ref. \cite{Aharony_77}. The first-order nature of the 
transition for negative (small) $v_0$ and large $N$ has also been 
confirmed in Ref. \cite{Wallace-73}.
\bibitem{footnote3}
The equation of state for the cubic-symmetric critical theory
has been calculated in Ref.~\cite{Aharony-74} to $O(\epsilon)$
in the framework of the field-theoretic $\epsilon$-expansion,
and in Ref.~\cite{B-S-92} to three loops in the 
fixed-dimension expansion for $N=0$.
\bibitem{footnote4}
For $N\ge 3$ one can use standard perturbation theory in the 
$N$-vector model to show that the cubic interaction is relevant.
One can also compute the crossover exponent in a $2+\epsilon$ expansion
\cite{Pelcovits-Nelson_76}.
\bibitem{footnote5} 
Note that in the $\epsilon$-expansion
one has expansions for the critical exponents at each fixed point.
Therefore,
the relations $\eta = \eta_I$ and  $\nu = \nu_I/(1-\alpha_I)$
must be satisfied order by order.
It can be checked that the five-loop series of Ref.~\cite{K-S-95}
satisfy these relations exactly.
\bibitem{footnoteVarnashev}
We can also compare the results for $\ub^*$ and $\vb^*$, noting that 
$\ub^* = u_c$ and $\vb^* = 9 v_c/(N + 8)$, where 
$u_c$ and $v_c$ are the results of Ref. \cite{Varnashev-99}.
The author finds: $\ub^* = 1.343(11)$, $\vb^* = 0.067(16)$ for $N=3$;
          $\ub^* = 0.906$, $\vb^* = 0.613$ for $N=4$;
          $\ub^* = 0.454$, $\vb^* = 1.129$ for $N=8$.
\end{references}
\end{document}